\documentclass{ieeeaccess}
\usepackage{cite}
\usepackage{amsmath,amssymb,amsfonts}

\usepackage{graphicx}
\usepackage{subfig}
\usepackage{textcomp}
\def\BibTeX{{\rm B\kern-.05em{\sc i\kern-.025em b}\kern-.08em
    T\kern-.1667em\lower.7ex\hbox{E}\kern-.125emX}}
    
\usepackage[table,xcdraw]{xcolor}
\usepackage{url}
\usepackage{multirow}
\usepackage{booktabs}
\usepackage{balance}
\usepackage{caption}
\usepackage{commath}
\usepackage{authblk}
\usepackage{ulem}  
\usepackage{array}

\usepackage{algorithm} 
\usepackage{algpseudocode}

\begin{document}
\doi{**xx.xxxx/xxxxxx**}
\title{A CNN-LSTM-based Fusion Separation Deep Neural Network for 6G Ultra-Massive MIMO Hybrid Beamforming}
\author{\uppercase{Rafid Umayer Murshed}\authorrefmark{1}, \uppercase{Zulqarnain Bin Ashraf}\authorrefmark{1},
\uppercase{Abu Horaira Hridhon}\authorrefmark{1},
\uppercase{Kumudu Munasinghe}\authorrefmark{2}, \uppercase{Abbas Jamalipour}\authorrefmark{3}
 AND \uppercase{Md. Farhad Hossain} \authorrefmark{1} 
}
\address[1]{Department of Electrical and Electronic Engineering, Bangladesh University of Engineering and Technology (BUET), Dhaka - 1205, Bangladesh.}
\address[2]{School of IT and Systems,
University of Canberra,
Canberra, Australia.}
\address[3]{School of Electrical \& Information Engineering,The University of Sydney, Australia.}


\markboth
{Rafid \headeretal: A CNN-LSTM-based Fusion Separation Network for 6G THz Hybrid Beamforming}
{Rafid \headeretal: A CNN-LSTM-based Fusion Separation Network for 6G THz Hybrid Beamforming}

\corresp{Corresponding author: Rafid Umayer Murshed(e-mail: rafid.buet.eee16@gmail.com).}

\begin{abstract}
In the sixth-generation (6G) cellular networks, hybrid beamforming would be a real-time optimization problem that is becoming progressively more challenging. Although numerical computation-based iterative methods such as the minimal mean square error (MMSE) and the alternative manifold-optimization (Alt-Min) can already attain near-optimal performance, their computational cost renders them unsuitable for real-time applications. However, recent studies have demonstrated that machine learning techniques like deep neural networks (DNN) can learn the mapping done by those algorithms between channel state information (CSI) and near-optimal resource allocation, and then approximate this mapping in near real-time. In light of this, we investigate various DNN architectures for beamforming challenges in the terahertz (THz) band for ultra-massive multiple-input multiple-output (UM-MIMO) and explore their contextual mathematical modeling. Specifically, we design a sophisticated 1D convolutional neural network and long short-term memory (1D CNN-LSTM) based fusion-separation scheme, which can approach the performance of the Alt-Min algorithm in terms of spectral efficiency (SE) and, at the same time, use significantly less computational effort. Simulation results indicate that the proposed system can attain almost the same level of SE as that of the numerical iterative algorithms, while incurring a substantial reduction in computational cost. Our DNN-based approach also exhibits exceptional adaptability to diverse network setups and high scalability. Although the current model only addresses the fully connected hybrid architecture, our approach can also be expanded to address a variety of other network topologies.
\end{abstract}


\titlepgskip=-15pt

\maketitle

\section{Introduction}
\label{sec:introduction}
\PARstart{W}{ith} daily increase in data traffic requirements ranging from mission-critical to massive machine connectivity, optimism for the sixth generation (6G) cellular network is growing exponentially. In response to the dramatic increase in the use of smartphones and related technologies, as well as the ongoing development of autonomous vehicles and Internet of Things (IoT) devices, the wireless industry has set extremely aggressive performance targets for the 6G systems. These targets will incorporate a peak data rate of at least 1 Tb/s, which is 100 times that of 5G, an over-the-air latency of 10–100 $\mu$s, and high mobility (1,000 km/h) \cite{zhang20206g}. Although specific requirements will vary according to deployment scenarios, it is safe to conclude that 6G will deliver a far more sophisticated user experience than current wireless technologies. Obtaining the desired performance levels often requires addressing complicated optimization problems extremely fast, such as resource allocation, beamforming, precoding, and scheduling.
The intriguing question now is how such enormous gains will be accomplished. Three main genres of improvement can be considered as listed below. 
\begin{itemize}
    \item Utilization of additional resources (e.g., gaining access to a greater chunk of the spectrum, including hitherto untapped frequency bands), 
    \item Enhancement of the efficiency of utilization of those resources using a variety of approaches (e.g., hybrid beamforming), and
    \item Virtualization of the network to the greatest extent possible in order to minimize cost, flexible deployment and better scalability (e.g., network slicing).
\end{itemize}
\par Although research has been going on in all these genres, increased utilization of existing resources has seen significant developments in the past couple of years. One of the most interesting prospects being explored for 6G deployment is the utilization of the THz band \cite{9585685}.

\par It is well-known that the huge accessible bandwidths at THz frequencies are accompanied by substantial propagation losses and power constraints, resulting in minimal communication distances. Several strategies have been proposed to combat this predicament, among which ultra-massive multiple-input multiple-output (UM-MIMO) beamforming seems to be the most promising \cite{8765243}. Furthermore, multiple antenna communication, a critical component of next-generation wireless technology, may attain higher data rates than those achieved by single antenna systems. The use of beamforming is essential for obtaining such high rates. These beamforming systems with many antenna elements can significantly enhance the spectral and energy efficiency at THz frequencies by focusing narrow and high-gain beams on a small region, thereby increasing the overall data rate and channel capacity via spatial multiplexing. Beamforming can primarily be classified into three types: analog, digital and hybrid. In analog beamforming, phase shifters are used in the RF domain at transmitters to send the same signal directed towards a specific receiver. These signals are combined at the receiver and hence coverage is increased. However, in digital beamforming, each transmitter is equipped with its own RF chain and multiple independent beams are generated, which significantly increases the data rate. This increase in data rate transpires at the cost of extremely high power consumption and hardware costs. On the other hand, hybrid beamforming (HBF) sacrifices a little accuracy to reduce power consumption and hardware costs significantly. Here, high-dimensional analog beamforming is combined with lower-dimensional digital baseband precoding. Due to its many advantages, hybrid beamforming has become quite popular in research and industry.

Existing beamforming algorithms rely on numerical iteration-based complex computations, which impede real-time resource allocation and generate a large amount of computational overhead, increasing both latency and costs \cite{HBF_cost}. Furthermore, these algorithms are often valid only for specific network configurations. With the enormous number of antennas required by THz UM-MIMO systems, the computational costs induced by these numerical iteration-based algorithms only get higher and higher. Furthermore, the maximum ratio (MR) and minimal mean square error (MMSE) algorithms utilized in MIMO receivers are not tuned to reduce computational complexity and communication latency. Consequently, there is an incentive to discover a different approach to beamforming problems that is simple, efficient and adaptable. Various algorithms have also been put forward to reduce power consumption through hybrid beamforming \cite{sohrabi2016hybrid}, which achieve sum rates comparable to those of fully digital beamformers. However, each algorithm developed thus far consumes a significant number of computational resources \cite{shi2011iteratively, sohrabi2016hybrid, el2014spatially}. Hence, they are impractical for real-time use. Meanwhile, the superior performance of deep neural networks (DNNs) on a range of inference and regression tasks has resulted in significant investment in research, development, and cloud infrastructure deployment for training and running DNNs. While the training phase of DNNs might be pretty time-consuming, the inference phase follows a straightforward deterministic execution model. Hence, DNNs are becoming increasingly more appealing for executing real-time inference tasks.

While substantial advances in the application of DL to wireless networks have been made \cite{9505313,9627726,9878155,9733558}, the majority of research in this domain is data-driven and does not make use of the most recent innovations in DL architectures and algorithms. This paper proposes a deep learning-based beamforming scheme with a specially designed fusion-separation network based upon a 1D convolutional neural network and long short-term memory (LSTM) for UM-MIMO beamforming at THz. This work combines domain-specific mathematical modeling with data-centric machine learning (ML) models to create a more sophisticated architecture with improved computational efficiency and robustness. We not only employ a data-driven strategy here but also include domain-specific knowledge and the most advanced ML techniques and architectures. We employ supervised learning to discover end-to-end mappings for MO-Alt-Min. Extensive simulations using synthetic data are conducted to validate our proposed approach.

\subsection{Related Works}
\par The design of the beamforming matrix is constrained by the expensive mmWave radio-frequency (RF) chains. Traditional full-digital beamformers need to connect a RF chain for each antenna element, which imposes intolerant power consumption and hardware cost and is no longer suitable for THz UM-MIMO  systems. Analog RF beamforming schemes, implemented using analog circuitry introduced in \cite{venkateswaran2010analog,chen2011multi,tsang2011coding,hur2013millimeter}, commonly employ analog phase shifters, which limit the beamformer's components to having a constant modulus. As a result, analog beamforming performs poorly in comparison to completely digital beamforming methods. To solve this problem, a HBF architecture was proposed in \cite{el2014spatially}. It replaces the fully digital
beamformer with a low-dimensional digital precoder followed by a high-dimensional analog precoder. 
\par To date,  most research focuses on two HBF structures, namely, fully-connected \cite{el2014spatially} and partially-connected architecture \cite{gao2016energy}. Authors in \cite{lin2016energy} studied fully connected and partially connected structures and demonstrated that the fully connected structure has better spectral and energy efficiency than the subconnected structure when insertion loss is included. Many topologies have been thoroughly researched to overcome the limitation of the number of RF chains. To obtain feasible hybrid beamformers and combiners, some numerical algorithms,
such as orthogonal matching pursuit (OMP) \cite{el2014spatially}, Karush-Kuhn-Tucker (KKT) based \cite{sohrabi2016hybrid}, and manifold optimization (MO) based \cite{yu2016alternating} algorithms, were proposed. The hybrid precoder design was approached in \cite{yu2016alternating} as a matrix factorization issue, which proposed efficient alternating minimization (Alt-Min) methods for the fully-connected and partially-connected hybrid precoding structures, respectively. In particular, an Alt-Min technique based on MO was suggested for the fully connected structure to approach the performance of the highly complex fully digital precoder. Thus, by imposing an orthogonal restriction on the digital precoder, a low-complexity Alt-Min algorithm was subsequently proposed. It, however, significantly reduced the SE. Therefore, these algorithms either do not achieve optimal performance in terms of SE or they generate a massive computational burden.

\par To this end, there has been much research on the implementation of beamforming in designing efficient and resilient MIMO systems \cite{yang2018digital,hassan2019edge,dutta2019case}. Existing hybrid beamforming systems frequently assume that infinite resolution phase shifters will be used to construct analog beamformers \cite{zhu2016novel}. However, the components necessary to construct precise phase shifters can be rather costly \cite{montori2010design,el2020large}. In practice, low-resolution phase shifters with a lower cost are frequently utilized. The simplest method for designing beamformers with limited resolution phase shifters is to build the RF beamformer with infinite resolution first and then to quantize the value of each phase shifter to a finite set \cite{liang2014low}. However, this approach is not useful in systems with low-resolution phase shifters \cite{sohrabi2015hybrid}. A specialized model is necessary for channels with enormous transmitting or receiving arrays or surfaces. However, as characteristics, such as path loss and angle of arrival (AoA), cannot be assumed constant between antennas, the channel model becomes non-stationary in space \cite{de2020non}. Another method for restricting the number of RF chains is to use simple analog switches to achieve antenna subset selection \cite{sanayei2004antenna,molisch2005capacity,sudarshan2006channel}. They cannot, however, achieve complete diversity gain in correlated channels since the antenna selection strategy uses just a subset of channels \cite{molisch2003reduced,molisch2004fft}.

\par Authors in \cite{el2014spatially} studied mmWave systems with massive antenna arrays. Precoding/combining was formulated as a sparse reconstruction task using mmWave channels' spatial structure. Using basis pursuit, algorithms were created that approximate optimal unconstrained precoders and combiners with low-cost RF hardware. Foad Sohrabi et al. \cite{sohrabi2016hybrid} proposed a hybrid beamforming design with a low-dimensional digital beamformer and an RF beamformer employing analog phase shifters. An architecture with fewer RF chains can approach the performance of a fully digital one. If the number of RF chains is double of the number of data streams, the hybrid beamforming structure can implement any completely digital beamformer regardless of antenna elements. In scenarios with fewer RF chains, the hybrid beamforming design problem was examined for both a point-to-point MIMO and a downlink MU-MISO system. However, these proposed systems do not take into account the computational complexity that would further aggravate in THz UM-MIMO.

\par To achieve near-optimal performance while reducing the long time consumption incurred by the conventional numerical algorithms, some DL based algorithms were proposed in \cite{elbir2019deep,elbir2019cnn,9815247,9353271}.
Authors in \cite{elbir2019deep} introduced a DL approach for joint channel estimation and hybrid beamforming for frequency selective, wideband mm-Wave systems. In particular, a massive MIMO orthogonal frequency division multiplexing (MIMO-OFDM) system was considered, and three different DL frameworks comprising convolutional neural networks (CNNs) were proposed, which accepted the received pilot signal as input and yielded the hybrid
beamformers at the output. Paper \cite{elbir2019cnn} further proposed a CNN framework for the joint design of precoder and combiners. This network accepts the input of a channel matrix and gives the output of analog and baseband beamformers. The proposed CNN framework does not require the knowledge of steering vectors of array responses, and it provides higher performance in capacity as compared to the conventional greedy- and optimization-based algorithms. However, as THz UM-MIMO channel matrices are extremely large in dimensions, the computational overhead generated would be too high for real-time implementation. These DL methods require a massive amount of training data in advance, and the dimension of their input data is quite high for UM-MIMO systems. The larger the input size of a neural network, the more computationally intensive it is. In some cases, the training data is very difficult to obtain. When the transmission environment or the system configuration changes, new training data is needed and the neural network needs to be retrained. In most cases, the entire architecture of the neural network has to be remodeled.

\begin{table}[t]
\caption{Summary of Important Abbreviations}
\label{abbreviations}
\centering
{
\begin{tabular}{|l|l|}
\hline
\textbf{Abbreviation} & \textbf{Definition}\\ [0.5ex] \hline
Alt-Min & Alternating Minimization\\[0.25ex]
\hline
CNN & Convolutional Neural Network\\[0.25ex] \hline
CSI & Channel State Information\\[0.25ex] \hline
DL & Deep Learning\\[0.25ex] \hline
DNN & Deep Neural Network\\[0.25ex] \hline
FPS-Alt-Min & Alt-Min algorithm using Fixed Phase Shifter\\[0.25ex] \hline
GMM & Gaussian Mixture Model\\[0.25ex] \hline
HBF & Hybrid Beamforming\\[0.25ex] \hline
IoT & Internet of Things\\[0.25ex] \hline
LSTM & Long Short-Term Memory\\[0.25ex]\hline
ML & Machine Learning\\[0.25ex]\hline
MMSE & Minimum Mean Squared Error\\[0.25ex]\hline
mmWave & Millimeter-Wave\\[0.25ex]\hline
MO & Manifold Optimization\\[0.25ex]\hline
MR & Maximum Ratio\\[0.25ex]\hline
OMP &  Orthogonal Matching Pursuit\\[0.25ex]\hline
PE-Alt-Min & Alt-Min algorithm using phase extraction\\[0.25ex]\hline
QoS & Quality of Service\\[0.25ex]\hline
RF & Radio Frequency\\[0.25ex]\hline
SE & Spectral Efficieny\\[0.25ex]\hline
SER & Symbol Error Rate\\[0.25ex]\hline
THz & Terahertz\\[0.25ex]\hline
ULA & Uniform Linear Array\\[0.25ex]\hline
URLLC & Ultra-Reliable Low Latency Communications\\[0.25ex]\hline
UM-MIMO & Ultra-Massive MIMO\\[0.25ex]\hline

\end{tabular}
}
\end{table}

\subsection{Contributions and Organization}
This paper presents a supervised learning-based hybrid beamforming scheme for point-to-point UM-MIMO. First, training data for the neural network is generated using OMP and Alt-Min algorithms. The network is then trained offline using synthetic data and deployed online. The main contributions of this work are summarized as follows:
\begin{itemize}
    \item We design a novel fusion-separation network to perform transmit and receive beamforming with reduced computational overhead. A single transmitter and receiver communicating at THz frequencies in a UM-MIMO system with multiple data streams is considered. We combine the sequence modeling capabilities of LSTM with the feature extraction capabilities of 1D-CNN to design a highly sophisticated neural network architecture. This beamforming scheme does not require the knowledge of steering vectors of array responses.
    \item We develop a mathematical modeling framework for the seamless simultaneous integration of complex matrices' real and imaginary components into a neural network's input layer. This allows us to reduce the complexity of the whole system and make it more efficient and lightweight.
    \item We analyze the performance of the novel fusion-separation network in terms of the achievable SE, varying RF chains, and computational overhead. The performance of the proposed approach is compared to that of the well-known Alt-Min algorithms. We also investigate the trade-off between model size and performance by creating three models of the same architecture but with different numbers of parameters.
\end{itemize}

The remainder of the paper is organized as follows. First, we delineate the system model and formulate the problem in Section II. Then, in Section III, we present the mathematical modeling of the proposed approach before delving into the detailed description of the novel fusion-separation network architecture in Section IV. Section V describes the simulation setup, while Section VI renders the simulation results accompanied by the pertinent analysis. Lastly, we conclude the paper with suggestions for future work in Section VII.

Regarding notation, scalars, matrices and vectors are represented in lower case, bold upper and bold lower cases, respectively. Scalar
norms, vector $L_2$ norms, Frobenius norms and pseudo-inverse are denoted by $|.|$, $||.||$, $||.||_F$ and $\dagger$, respectively. For any general matrix or vector operator \textbf{x}, \textbf{x\textsuperscript{T}} and \textbf{x\textsuperscript{*}} represent the transpose and conjugate transpose matrices, respectively. E[.], $\mathbb{C}$ and $\mathbb{N}$ denote the expected value, the set of the complex and natural numbers, respectively. Table \ref{abbreviations} and Table \ref{notations} present the summary of important abbreviations and the system model notations, respectively.
\begin{table}[t]
\caption{Summary of System Model Notations}
\label{notations}
\centering
{
\begin{tabular}{|l|l|}
\hline
\textbf{Notation} & \textbf{Definition}\\ [0.5ex] \hline
\textbf{a\textsubscript t, a\textsubscript r} & Transmit and receive SA steering vectors\\[0.25ex]
\hline
\textbf{b} & Binary symbol vector\\[0.25ex] \hline
C(A) & Column space of A\\[0.25ex] \hline
\textbf{e} & Error vector\\[0.25ex] \hline
\textit{f} & Carrier frequency\\[0.25ex] \hline
\textbf{F}\textsubscript{opt} & Optimal fully digital precoder matrix\\[0.25ex] \hline
$\textbf{F_{RF}}$, $\textbf{F_{BB}}$ & Analog and digital transmit beamforming matrices\\[0.25ex] \hline
\textbf{H} & Channel matrix\\[0.25ex]\hline
\textit{N\textsubscript{cl}}, \textit{N\textsubscript{ray}}  & Number of clusters and number of rays in each cluster\\[0.25ex]\hline
\textit{N\textsubscript{p}} & Number of propagation paths\\[0.25ex]\hline
\textit{N\textsubscript{s}} & Number of data-streams\\[0.25ex]\hline
\textit{N\textsubscript{t}}, \textit{N\textsubscript{r}} &  Number of antennas at the transmitter and the receiver\\[0.25ex]\hline
\textit{N\rlap{\textsubscript{RF}}\textsuperscript{t}} \;, \textit{N\rlap{\textsubscript{RF}}\textsuperscript{r}} & Number of RF chains at the transmitter and the receiver\\[0.25ex]\hline
P &  Angular power delay profile \\[0.25ex]\hline
\textbf{s} & Symbol vector\\[0.25ex]\hline
$\textbf{W_{RF}}$, $\textbf{W_{BB}}$ & Analog and digital beamforming matrices at receiver end\\[0.25ex]\hline
$\textbf{W_{t}}$, $\textbf{W_{r}}$ & Precoding and combining matrices\\[0.25ex]\hline
\textbf{x} & Transmitted signal\\[0.25ex]\hline
$\alpha$ & Path gain\\[0.25ex]\hline
$\beta\textsubscript{n}$ &  Complex gain of the nth path \\[0.25ex]\hline
$\rho$ & Transmit power\\[0.25ex]\hline
$\phi\textsubscript{r}$, $\theta\textsubscript{r}$ & Receive angles of departure and arrival\\[0.25ex]\hline
$\phi\textsubscript{t}$, $\theta\textsubscript{t}$ & Transmit angles of departure and arrival\\[0.25ex]\hline
\end{tabular}
}
\end{table}

\section{System Model and Problem Formulation}\label{system model}
In this section, we describe the adopted network and channel models. We then move on to formulate the basic beamforming problem.
\begin{figure*}[! h]
\centering
    \includegraphics[width=2\columnwidth,height=90mm]{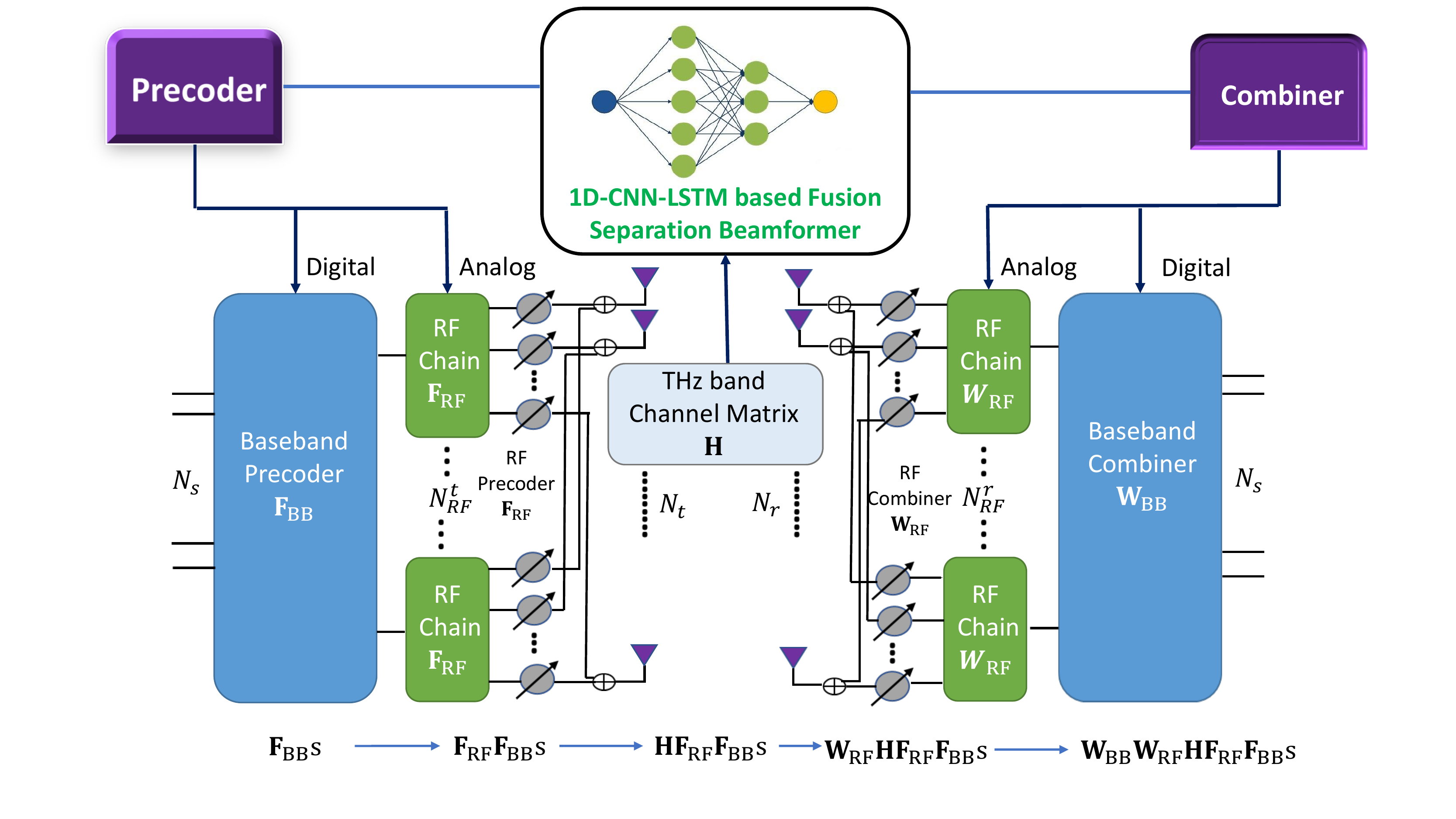}
    \caption{System architecture for 1D CNN-LSTM based UM-MIMO transceiver with hybrid beamforming.}
    \vspace{-2mm}
    \label{networkmodel}
\end{figure*}
\subsection{Network Model}

\par We consider a single-user UM-MIMO network, where both the base station (BS) and the user employ a uniform linear array (ULA) of multiple antennas. We assume that there are $N_{t}$ antennas at the transmitter and $N_{r}$ antennas at the receiver as shown in Figure \ref{networkmodel}. There are $N^t_{RF}$ and $N^r_{RF}$ number of RF chains at the transmitter and the receiver, respectively. We also consider this as a point-to-point multi-stream network with $N_{s}$ number of data-streams. Here,
$N_{s}\le N^t_{RF}\le N_{t}$ and $N_{s}\le N^r_{RF}\le N_{r}$. Due to the limited number of transmit/receive RF chains, it is not possible to execute completely digital beamforming, which requires one dedicated RF chain per antenna element. As depicted in Figure \ref{networkmodel}, we instead explore a two-stage hybrid digital and analog beamforming architecture at the BS and user terminal. Now, the transmitted signal can be given as below
\begin{equation}
\label{equation1}
  \textbf{x}= \textbf{F}\textsubscript{RF}\textbf{F}\textsubscript{BB}\textbf{s},
\end{equation}
where $\textbf{F}\textsubscript{BB} \in \mathbb{C}^{N^t_{RF}\times N_{s}}$ is the $N^t_{RF} \times N_s$ digital baseband precoder, $\textbf{F}\textsubscript{RF} \in \mathbb{C}^{N_{t} \times N^t_{RF}}$ is the $N_t \times N^t_{RF}$ analog RF precoder, and $ \textbf{s} \in \mathbb{C}^{N_{s}}$ is the symbol vector with 
\begin{equation}
   E[\textbf{s}\textbf{s}^H]= (\frac{1}{N_s})\textbf{I}_{\textbf{N_s}}.
\end{equation}
\par Now, the received signal \textbf{r} after decoding can be expressed as
\begin{equation}
\label{equation2}
  \textbf{r}= \sqrt{\rho}\textbf{W}\textsubscript{RF}^H\textbf{W\textsubscript{BB}}^H\textbf{H}\textbf{x} + \textbf{W\textsubscript{RF}}^H\textbf{W\textsubscript{BB}}^H\textbf{n},
\end{equation}
where $\textbf{W\textsubscript{RF}} \in \mathbb{C}^{N_{r} \times N^r_{RF}}$ and $\textbf{W\textsubscript{BB}} \in \mathbb{C}^{N^r_{RF}\times N_{s}}$ are respectively the analog and digital beamforming matrices at the receiver end, $\rho$ is the transmit power, and $\textbf{n} \sim \mathcal{C}\mathcal{N}(0, \sigma^2\textbf{I}_{N_r})$ denotes the additive white Gaussian noise (AWGN) vector.

\subsection{Channel Model}

In this research, the Saleh-Valenzuela (S-V) model with a THz-band modification is employed to develop the channel model. The S-V model is a cluster-based, statistical channel model that connects clustering phenomena to stochastic angles of departure and arrival for each beam.  
Furthermore, we assume perfect CSI at the transmitter and the receiver.
The channel matrix $\textbf{H}\in \mathbb{C}^{N^t\times N^r}$ can be given by
\begin{equation}
   \textbf{H} = 
   \sqrt{\frac{N_t\;N_r}{N_{cl} \; N_{ray}}}
    \;\sum_{i=1}^{N_{cl}}\;\sum_{l=1}^{N_{ray}}\alpha_{il}\textbf{a_r}(\phi^r_{il},\theta^r_{il})\;\textbf{a_t}(\phi^t_{il},\theta^t_{il})^H,
    \end{equation}
where \textit{N\textsubscript{cl}} and \textit{N\textsubscript{ray}} represent the number of clusters and the number of rays in each cluster, respectively. \textit{$\alpha \textsubscript{il}$} represents the gain of the $l^{th}$ ray in the $i^{th}$ propagation cluster. We assume that \textit{$\alpha \textsubscript{il}$} are independent and identically distributed according to the distribution $\mathcal{C}\mathcal{N}(0, \sigma^2_{\alpha,i})$ and $\sum_{i=1}^{N_{cl}}\sigma^2_{\alpha,i}= \hat{\gamma}$, which is the normalization factor to satisfy $\mathbb{E}[||\textbf{H}||^2_F]=N_tN_r$. Moreover, $\textbf{a_r}(\phi^r_{il},\theta^r_{il})$ and $\textbf{a_t}(\phi^t_{il},\theta^t_{il})$
denote the receive and transmit array response vectors,
where $\phi^r_{il}(\phi^t_{il})$ and $\theta^r_{il}(\theta^t_{il}))$ stand for azimuth and elevation angles of arrival and departure, respectively. In this study, the uniform square planar array (USPA) with $\sqrt{N}\times \sqrt{N}$ antenna elements is investigated. Therefore, the array response vector corresponding to the $l^{th}$ ray in the $i^{th}$ cluster can be written as
\begin{equation}
\begin{gathered}
\textbf{a}(\phi_{il},\theta_{il}) = \frac{1}{\sqrt{N}}[1,..., e^{j\frac{2 \pi}{\lambda}d(p\sin{\phi_{il}\sin{\theta_{il}}}\;+\;q\cos{\theta_{il}})},...,\\
   e^{j\frac{2 \pi}{\lambda}d((\sqrt{N}-1)\;\sin{\phi_{il}\sin{\theta_{il}}}\;+\;(\sqrt{N}-1)\cos{\theta_{il}})}]^T
\end{gathered},
   \end{equation}
where d and $\lambda$ represent the antenna spacing and signal wavelength respectively, and $0\le p<\sqrt{N}$ and $0\le q<\sqrt{N}$ represent the antenna indices in the 2D plane, respectively. This channel model will be utilized in simulations, but our beamformer architecture is also applicable to more general models.

\subsubsection{Azimuth $\And$ Elevation Angle:}
Denoting the angular power delay profile of each individual cluster as $P_i(\phi,\theta)_{cluster}$, the complete power delay profile can be computed as\cite{priebe2011aoa}
\begin{equation}
    P(\phi,\theta)= \sum_{i=1}^{N_{c}} P_i(\phi,\theta)_{cluster}.
\end{equation}
\par Furthermore, considering there is no correlation between angular power profiles, the angular power profile for each cluster can be rewritten as
\begin{equation}
    P_i(\phi,\theta)_{cluster} = P_i(\phi)_{cluster}\;P_i(\theta)_{cluster}.
\end{equation}
\par The azimuth and elevation angle of each ray is assumed to be independent \cite{840194}. If the azimuth angle of $i^{th}$ cluster is $\varphi_i^t$, and the azimuth angle of $j^{th}$ ray within the $i^{th}$ cluster is $\Phi_{ij}^t$, then the total azimuth angle is expressed as
\begin{equation}
    \phi_{ij}^t = \varphi_i^t+ \Phi_{ij}^t.
\end{equation}
\par Similarly, if the elevation angle of $i^{th}$ cluster is $\vartheta_i^t$, and elevation angle of $j^{th}$ ray within the $i^{th}$ cluster is $\Theta_{ij}^t$, then the total elevation angle is
\begin{equation}
    	\theta_{ij}^t = \vartheta_i^t + \Theta_{ij}^t,
\end{equation}
where $\Phi_{ij}^t$ and $\Theta_{ij}^t$ follow zero-mean second order Gaussian mixture model that can be expressed as \cite{7036065}

\begin{equation}
\begin{gathered}
    GMM(x) = \frac{a_1}{2\pi \sigma_1} e^{-\frac{1}{2}(\frac{x- \overline{x_1}}{\sigma_1})^2} +
    \frac{a_2}{2\pi \sigma_2} e^{-\frac{1}{2}(\frac{x- \overline{x_2}}{\sigma_1})^2},\\
    \end{gathered}
\end{equation}
where $\overline{x_1}, \overline{x_2} = 0$.
Azimuth angle $\varphi_i^t$ and elevation angle $\vartheta_i^t$ of each cluster follow the uniform distribution, where $\varphi_i^t \in (-\pi,\pi]$ and $\vartheta_i^t \in [-\frac{\pi}{2},\frac{\pi}{2}]$.

\subsection{Problem Formulation}
The beamforming problem can be segregated into two independent sub-problems, i.e., the precoder design and the decoder design. Their mathematical formulations are almost identical, aside from the former having an additional power constraint. The precoder design problem can be formulated as below
\begin{equation}\label{4.6}
\begin{gathered}
\min_{\textbf{F\textsubscript{RF}}\textbf{F\textsubscript{BB}}}  \norm{\textbf{F\textsubscript{opt}} - \textbf{F\textsubscript{RF}}\textbf{F\textsubscript{BB}}}_F
\\
s.t. \; |(\textbf{F\textsubscript{RF}})_{ij}|=1, \;   \norm{\textbf{F\textsubscript{RF}},\textbf{F\textsubscript{BB}}}^2_F=N_s.
\end{gathered}
\end{equation}

It has been proven that the above problem (\ref{4.6}) is an analogous formulation for maximizing SE \cite{8646553}. This can be fairly evident since the optimal hybrid precoders are closely comparable to the unconstrained optimal fully digital precoder. Due to the fact that the optimal fully digital precoder matrix \textbf{F}\textsubscript{opt} $\in \mathbb{R}^{N_{t} \times N_{s}}$ contains the eigen vectors of the channel matrix \textbf{H}, \textbf{F}\textsubscript{opt} corresponds to the first $N_{s}$ columns of \textbf{V}, where \textbf{V} is generated from the singular value decomposition (SVD) of the channel \textbf{H}, i.e. $\textbf{H} = \textbf{U}\;\sum \textbf{V\textsuperscript{H}}$.\\

 Once the beamforming matrices are calculated, SE is found by
 \begin{equation}
R= \log _{2}{|\textbf{I_M}+\frac{\rho}{N_s}\textbf{W_t}(\textbf{W^H_t}\textbf{W_t})^{-1}\textbf{W^H_t}\textbf{H}\textbf{V_t}\textbf{V^H_t}\textbf{H^H}|},
 \end{equation}
 where $\textbf{W\textsubscript{t}}=\textbf{W\textsubscript{RF}}\times \textbf{W\textsubscript{BB}}$, $\textbf{V_{t}}= \textbf{F\textsubscript{RF}}\times \textbf{F\textsubscript{BB}}$ and $\rho$ is the transmit power.
\section{Mathematical Modelling}
\begin{figure}[t]
    \includegraphics[width=\linewidth]{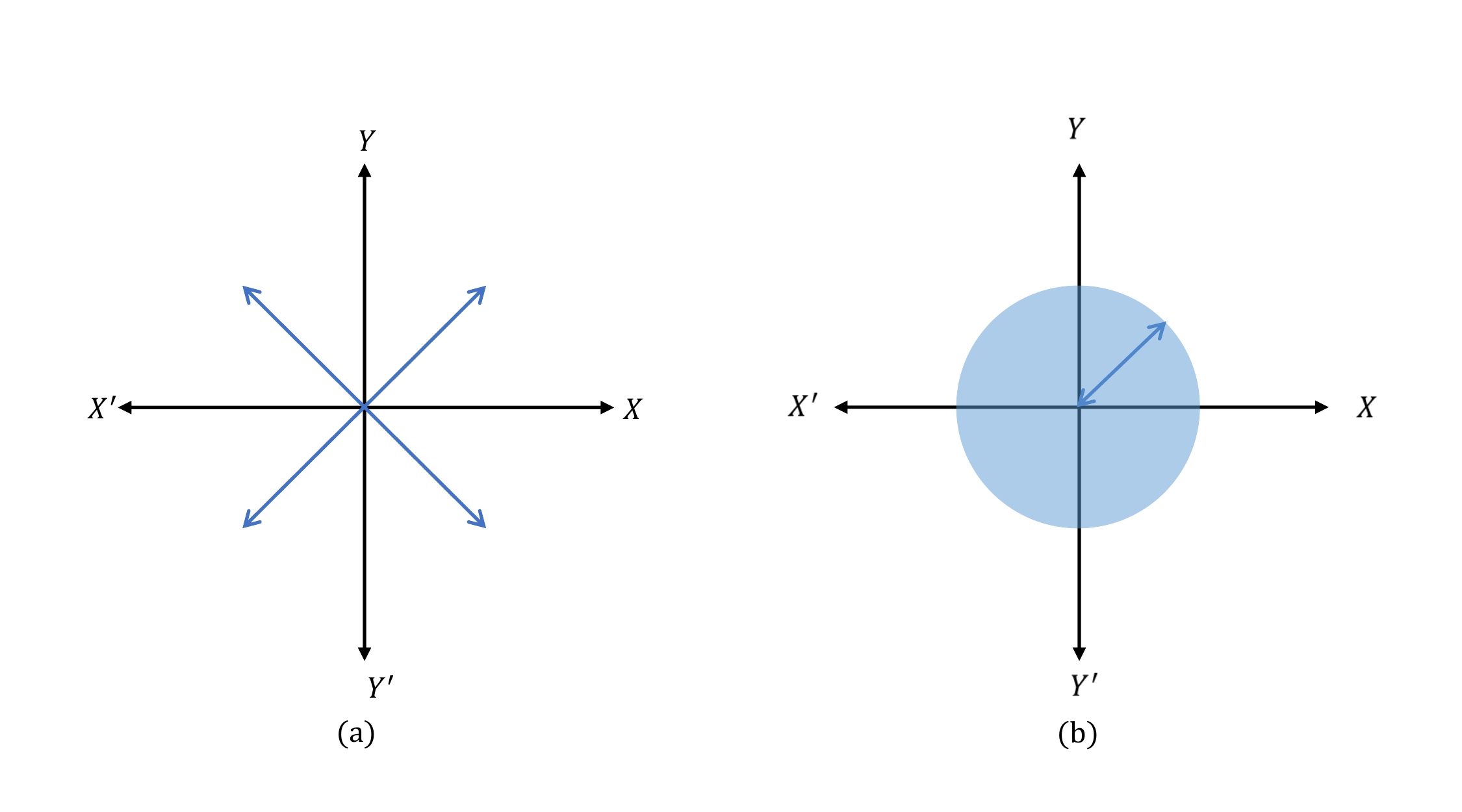}
    \caption{2D projection of beamforming matrices: (a) Possible directions in which \textbf{F}\textsubscript{RF}\textbf{F}\textsubscript{BB} can point, directions are limited due to constraints on \textbf{F}\textsubscript{RF}; (b) \textbf{F}\textsubscript{opt} can point in any direction. }
    \vspace{-5mm}
    \label{mathmodel1}
\end{figure}

\begin{figure*}[t]
\centering
    \includegraphics[scale=0.5]{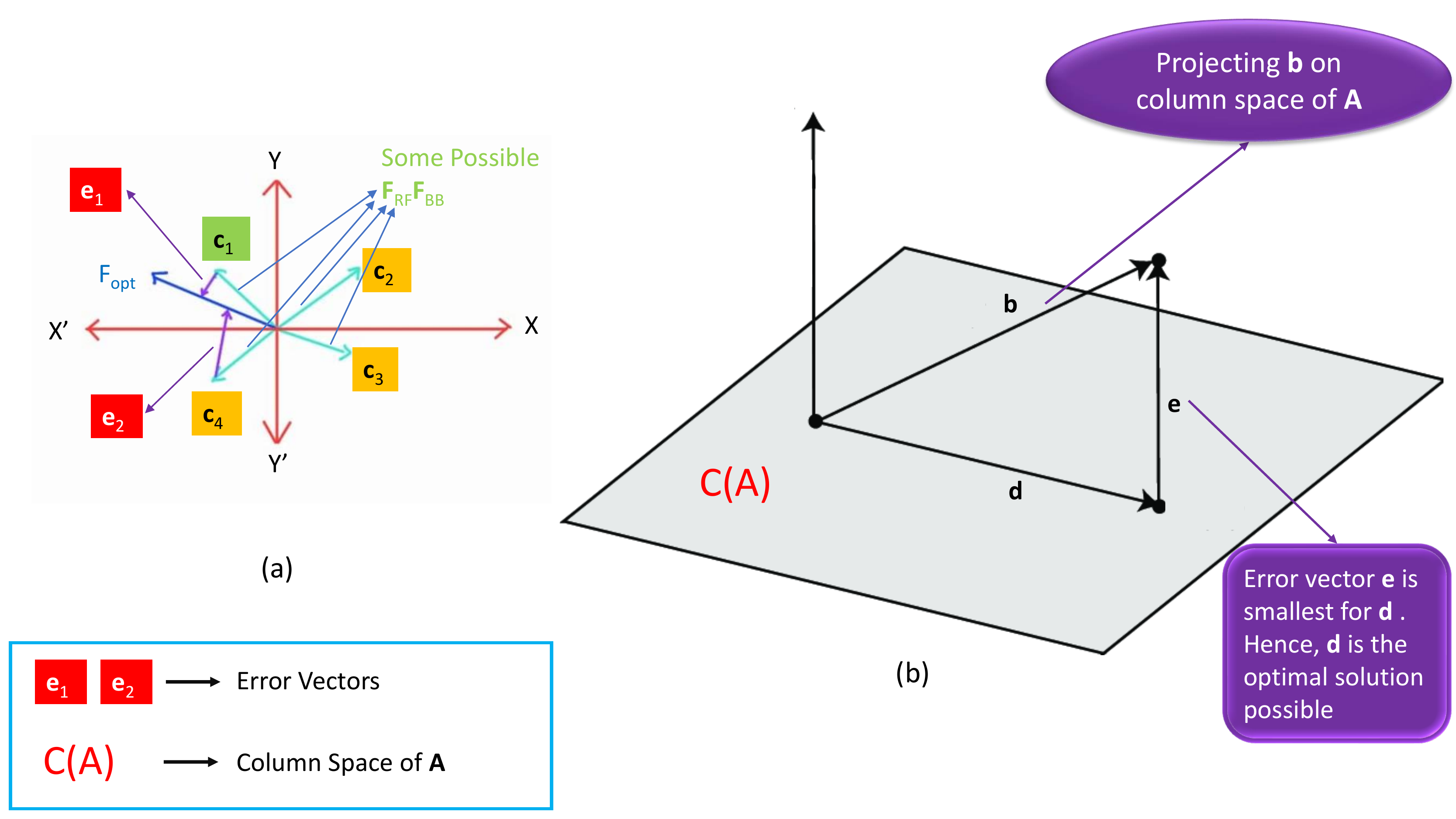}
    \caption{Hypothetical vector representation of beamforming matrix columns}
    \vspace{-2mm}
    \label{mathmodel2}
\end{figure*}

Before delving into the details of our proposed system, let’s take a quick look at how we mathematically model the problem and what our DNN's approach to solving it is. Let us first think of \textbf{F}\textsubscript{opt} and \textbf{F}\textsubscript{RF}\textbf{F}\textsubscript{BB} matrices as vectors in a very high-dimensional complex plane, which we hypothetically project onto a 2D real plane here for prompt analogy. Due to constraints on \textbf{F}\textsubscript{RF}, the vector of \textbf{F}\textsubscript{RF}\textbf{F}\textsubscript{BB} cannot point in all directions as shown in Figure \ref{mathmodel1}(a). 
However, \textbf{F}\textsubscript{opt} can point in any given direction as shown in Figure \ref{mathmodel1}(b). Since, \textbf{F}\textsubscript{RF}\textbf{F}\textsubscript{BB} cannot point in every direction like \textbf{F}\textsubscript{opt}, \textbf{F}\textsubscript{RF}\textbf{F}\textsubscript{BB} will not be equal to \textbf{F}\textsubscript{opt}  most of the time. Hence, there will always be an error vector \textbf{e} between these two vectors as shown in Figure \ref{mathmodel2}. Our goal is to minimize this error vector in a much higher-dimensional complex plane. In other words, we need to find the \textbf{F}\textsubscript{RF}\textbf{F}\textsubscript{BB} which is closest to \textbf{F}\textsubscript{opt}. In this analogy, from Figure \ref{mathmodel2} we observe that \textbf{c}\textsubscript{1} is the closest possible choice because its projection on \textbf{F}\textsubscript{opt} is the largest and hence its error \textbf{e}\textsubscript{1} is also the smallest.
We can also consider the beamforming problem (\ref{4.6}) as a system of matrix equations of the form \textbf{Ax} = \textbf{b}, which we can solve for an exact \textbf{x} only when \textbf{b} is in the column space of \textbf{A}. We can consider each column of \textbf{F}\textsubscript{BB} and \textbf{F}\textsubscript{opt} as vectors in $N^t_{RF}$ and $N_t$ dimensional space respectively. This can be expressed in matrix form as
\begin{equation}\label{bigAxeqb}
\textbf{A}\textbf{X}=\textbf{B},
\end{equation} 
where
\\
\\
\textbf{A} = $\begin{bmatrix}
A_{11} & ...& A_{1{N_{RF}}}
\\
A_{21} & ...& A_{2{N_{RF}}}
\\
.&.&.
\\
.&.&.
\\
.&.&.
\\
A_{N_t1} & ...& A_{N_t{N_{RF}}}
\end{bmatrix}$,
\textbf{X} = $\begin{bmatrix}
x_{11} & ...& x_{1{N_{s}}}
\\
.&.&.
\\
.&.&.
\\
.&.&.
\\
x_{N_{RF1}} & ...& x_{N_{RF}N_{s}}
\end{bmatrix}$
\\
\\
\\
and
$\textbf{B} = \begin{bmatrix}
b_{11} & ...& b_{1{N_{s}}}
\\
b_{21} & ...& b_{2{N_{s}}}
\\
.&.&.
\\
.&.&.
\\
.&.&.
\\
b_{N_t1} & ...& b_{N_t{N_{s}}}
\end{bmatrix}$.
\\
\\
  Here, matrix \textbf{A}, \textbf{B} and \textbf{X} corresponds to \textbf{F}\textsubscript{RF}, \textbf{F}\textsubscript{opt} and \textbf{F}\textsubscript{BB} respectively. We now resolve (\ref{bigAxeqb}) into $N_s$ equations of the form \textbf{Ax} = \textbf{b} which can be expressed as 
 \begin{equation}\label{resAxeqb}
\resizebox{1\hsize}{!}{
$\begin{bmatrix}
A_{11} & ...& A_{1{N_{RF}}}
\\
A_{21} & ...& A_{2{N_{RF}}}
\\
.&.&.
\\
.&.&.
\\
.&.&.
\\
A_{N_t1} & ...& A_{N_t{N_{RF}}}
\end{bmatrix} 
\times
\begin{bmatrix}
x_{11}
\\
.
\\
.
\\
.
\\
x_{N_{RF}1}
\end{bmatrix}
=
\begin{bmatrix}
b_{11}
\\
b_{21}
\\
.
\\
.
\\
.
\\
b_{N_t1}
\end{bmatrix}$}.
\end{equation}
 \par Here, (\ref{resAxeqb}) can only be solved for \textbf{x} when the vector \textbf{b} is in the column space of the matrix \textbf{A}. Unfortunately, due to the unit modulus constraint, our column space of \textbf{A} is severely limited. In simple words, the dimension C(\textbf{A}) will be much less than the dimension of \textbf{b}. Therefore, we must solve for \textbf{Ay} = \textbf{d}, where \textbf{d} will be the projection of \textbf{b} on the column space of \textbf{A}. Consequently, we can rewrite (\ref{resAxeqb}) as
 \begin{equation}\label{ayeqd}
\resizebox{1\hsize}{!}{
          $\begin{bmatrix}
A_{11} & ...& A_{1{N_{RF}}}
\\
A_{21} & ...& A_{2{N_{RF}}}
\\
.&.&.
\\
.&.&.
\\
.&.&.
\\
A_{N_t1} & ...& A_{N_t{N_{RF}}}
\end{bmatrix} 
\times
          \begin{bmatrix}
y_{11}
\\
.
\\
.
\\
.
\\
y_{N_{RF}1} 
\end{bmatrix}
\\\\
=
\begin{bmatrix}
d_{11}
\\
d_{21}
\\
.
\\
.
\\
.
\\
d_{N_t1}
\end{bmatrix}$}.
\end{equation}
\par Now problem (\ref{4.6}) involves solving \textbf{Ay} = \textbf{d} as given by (\ref{ayeqd}) for every column in \textbf{F}\textsubscript{BB}.
Below we show that (\ref{4.6}) essentially requires finding the best possible \textbf{d} for every \textbf{b} in \textbf{F}\textsubscript{opt} or in other words minimizing \textbf{e} for every \textbf{b} as depicted in Figure \ref{mathmodel2}(b).
\begin{itemize}
\item $\textbf{e} = \textbf{b}-\textbf{d}$ and minimizing \textbf{e} means minimizing |\textbf{b}-\textbf{d}|.
\item Again, $\textbf{d} = \textbf{A}\textbf{y}$. Hence, we must minimize $|\textbf{b}-\textbf{A}\textbf{y}|$ for every \textbf{b} and \textbf{y}.
\item But, all the \textbf{b}’s are the columns of \textbf{F}\textsubscript{opt}, all the \textbf{y}’s are the columns of the optimal \textbf{F}\textsubscript{BB} and \textbf{A} is the \textbf{F}\textsubscript{RF}.
\item Hence, minimizing $|\textbf{b}-\textbf{A}\textbf{y}|$ for every b is the same as minimizing the Frobenius norm of \textbf{F}\textsubscript{opt} - \textbf{F}\textsubscript{RF}\textbf{F}\textsubscript{BB}, which is required by (\ref{4.6}).
\end{itemize}

 \begin{figure*}[t]
    \includegraphics[height=100mm,width=\linewidth]{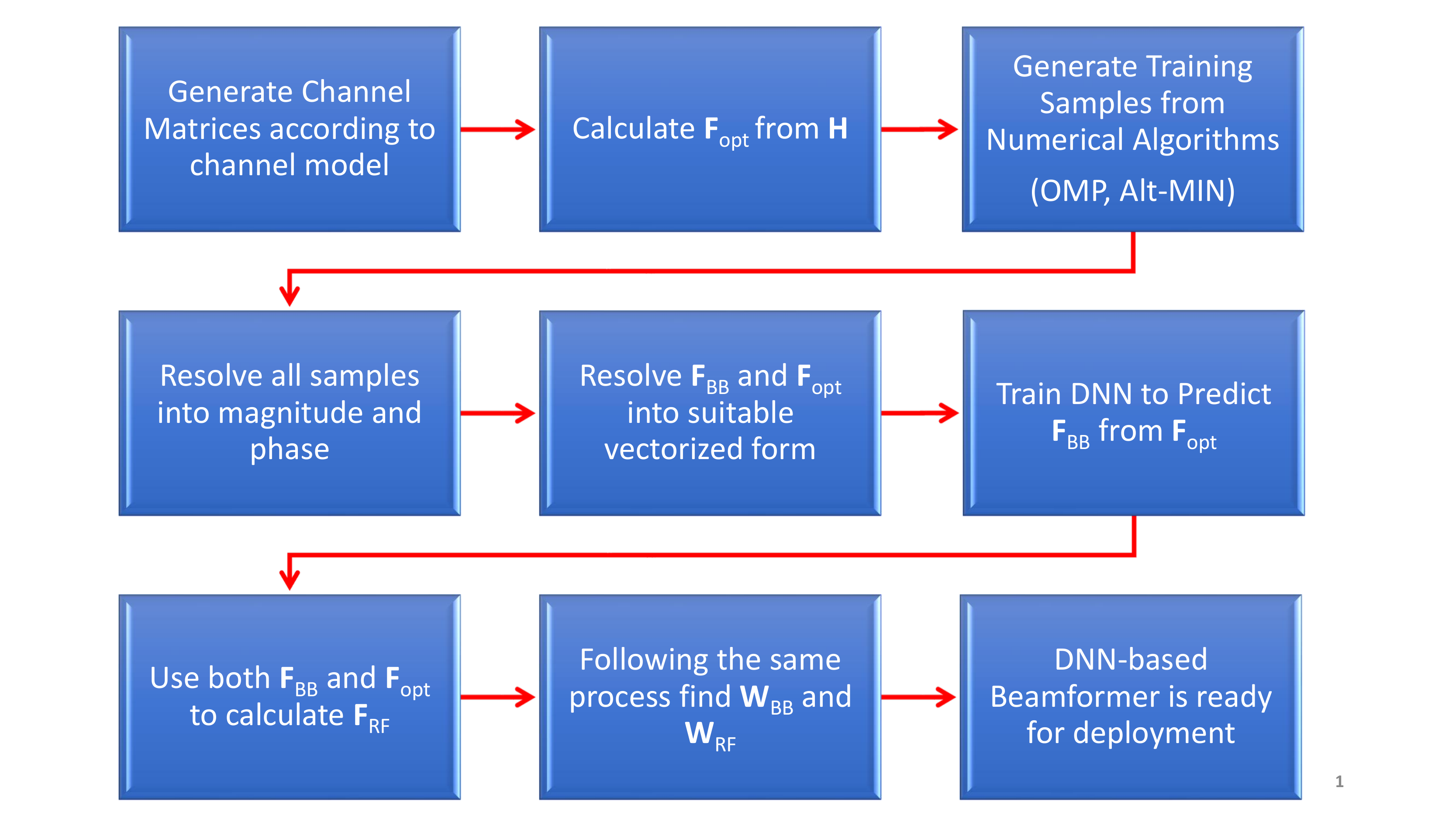}
    \caption{Workflow of the proposed hybrid beamforming system.}
    \vspace{-2mm}
    \label{workflow}
\end{figure*}

\section{Proposed DNN based Beamforming System}
\label{sec:proposeddnn}
This section presents the specifics of the proposed DNN system for hybrid beamforming. First, we describe the procedure by which simulation data for training, validation, and testing of the proposed system is created. The design and architecture of the neural network utilized to construct the beamforming matrices are then discussed. Then, we delineate how the input data and features travel through the network architecture to produce the desired outputs. Finally, we demonstrate how the beamforming matrices are computed using the neural network's outputs.
\subsection{Data Generation and Pre-processing}
Our proposed DNN actually solves the beam-forming problem by essentially solving the \textbf{Ay} = \textbf{d} problem as given in (\ref{ayeqd}). Each training sample consists of one column of \textbf{F}\textsubscript{BB}(=\textbf{y}) and the corresponding column from \textbf{F}\textsubscript{opt}(=\textbf{b}) as illustrated below
\begin{equation}
\begin{aligned}
&
\left.
\begin{bmatrix}
A_{11} & ...& A_{1{N_{RF}}}
\\
A_{21} & ...& A_{2{N_{RF}}}
\\
.&.&.
\\
.&.&.
\\
.&.&.
\\
A_{N_{t1}} & ...& A_{N_t{N_{RF}}}
\end{bmatrix}
\times
\begin{bmatrix}
y_{11} &...& y_{1{N_{s}}}
\\
.&.&.
\\
.&.&.
\\
.&.&.
\\
y_{N_{RF1}} & ...& y_{N_{RF}{N_{s}}}
\end{bmatrix}\right.\\
&\qquad\qquad
\left.
=
\right.
\begin{bmatrix}
b_{11} & ...& b_{1{N_{s}}}
\\
b_{21} & ...& b_{2{N_{s}}}
\\
.&.&.
\\
.&.&.
\\
.&.&.
\\
b_{N_{t1}} & ...& b_{N_t{N_{s}}}
\end{bmatrix}.
\end{aligned}
\end{equation}


 Here, the column [$y_{11}$, $y_{21}$,...., $y_{N_{RF1}}$]\textsuperscript{T} of \textbf{F}\textsubscript{BB} and corresponding column [$b_{11}$, $b_{21}$,...., $b_{N_{t1}}$]\textsuperscript{T} of \textbf{F}\textsubscript{opt} constitute one training sample. Now, we look at exactly how the ground truth of our DNN is generated. As we can see, we treat every column of \textbf{F}\textsubscript{opt} and \textbf{F}\textsubscript{BB} as an individual training sample. Our DNN learns by looking at thousands of possible pairs of \textbf{y} and \textbf{b}. We also observe that each column of \textbf{F}\textsubscript{opt} is related to only the corresponding column of \textbf{F}\textsubscript{BB}. 
 Our DNN learns to predict columns of \textbf{F}\textsubscript{BB} by looking at columns of \textbf{F}\textsubscript{opt}. Finally, \textbf{F}\textsubscript{RF} is calculated from \textbf{F}\textsubscript{opt} and \textbf{F}\textsubscript{BB} through pseudo-inverse.
\par In order to produce simulation data for training, validation and testing, we start by generating thousands of instances of the channel matrix \textbf{H} from the channel model described in section \ref{system model}. Then, corresponding to each instance of the channel matrix, we calculate the fully digital optimal beamforming matrix \textbf{F}\textsubscript{opt} from the SVD of \textbf{H}. We then utilize the existing numerical iteration-based algorithms such as the OMP and Alt-Min to find the beamforming matrices \textbf{F}\textsubscript{RF}, \textbf{F}\textsubscript{BB}, \textbf{W}\textsubscript{RF}, \textbf{W}\textsubscript{BB}. Since most neural networks are not configured to work with complex numbers, the complex numbers are resolved into their polar form. Furthermore, we split up the phase and magnitude of the complex matrices. Now, these matrices can be treated in the same way as ordinary real number matrices. Finally, following standard practice, the input data is normalized before being passed onto the network. Mathematically, the magnitude normalization can be expressed as
\begin{equation}
   v_M(i)=\frac{u_M(i)}{max(u_M)}
   .\end{equation}
\par Here, $u_M(i)$ denotes the magnitude of any element of the matrix
\textbf{u} $\in$ [\textbf{F}\textsubscript{opt},
 \textbf{F}\textsubscript{BB} and \textbf{W}\textsubscript{BB}], max(.) gives the global maximum and $v_M(i)$ denotes the normalized magnitude input that would be passed to the neural network. Similarly, the phase is also normalized and is given by
\begin{equation}
   v_P(i)=\frac{u_P(i)+\pi}{2\pi},
   \end{equation}
where $u_P(i)$ denotes the phase of any element of the matrix \textbf{u} $\in$ [\textbf{F}\textsubscript{opt}, \textbf{F}\textsubscript{BB} and \textbf{W}\textsubscript{BB}] and $v_P(i)$ denotes the normalized phase input that would be passed to the neural network. Moreover, according to the mathematical modelling presented in section III, we split up the matrices \textbf{F}\textsubscript{opt}, \textbf{F}\textsubscript{BB} and \textbf{W}\textsubscript{BB} into their constituent columns, that is we vectorize them. Finally, these resolved columns of \textbf{F}\textsubscript{opt} are passed onto the input layer of the neural network. This entire workflow is summarized in Figure 4.
\subsection{Architecture of the Proposed DNN}
\begin{figure*}[ht]
\centering
    \includegraphics[scale=0.5, width=\linewidth]{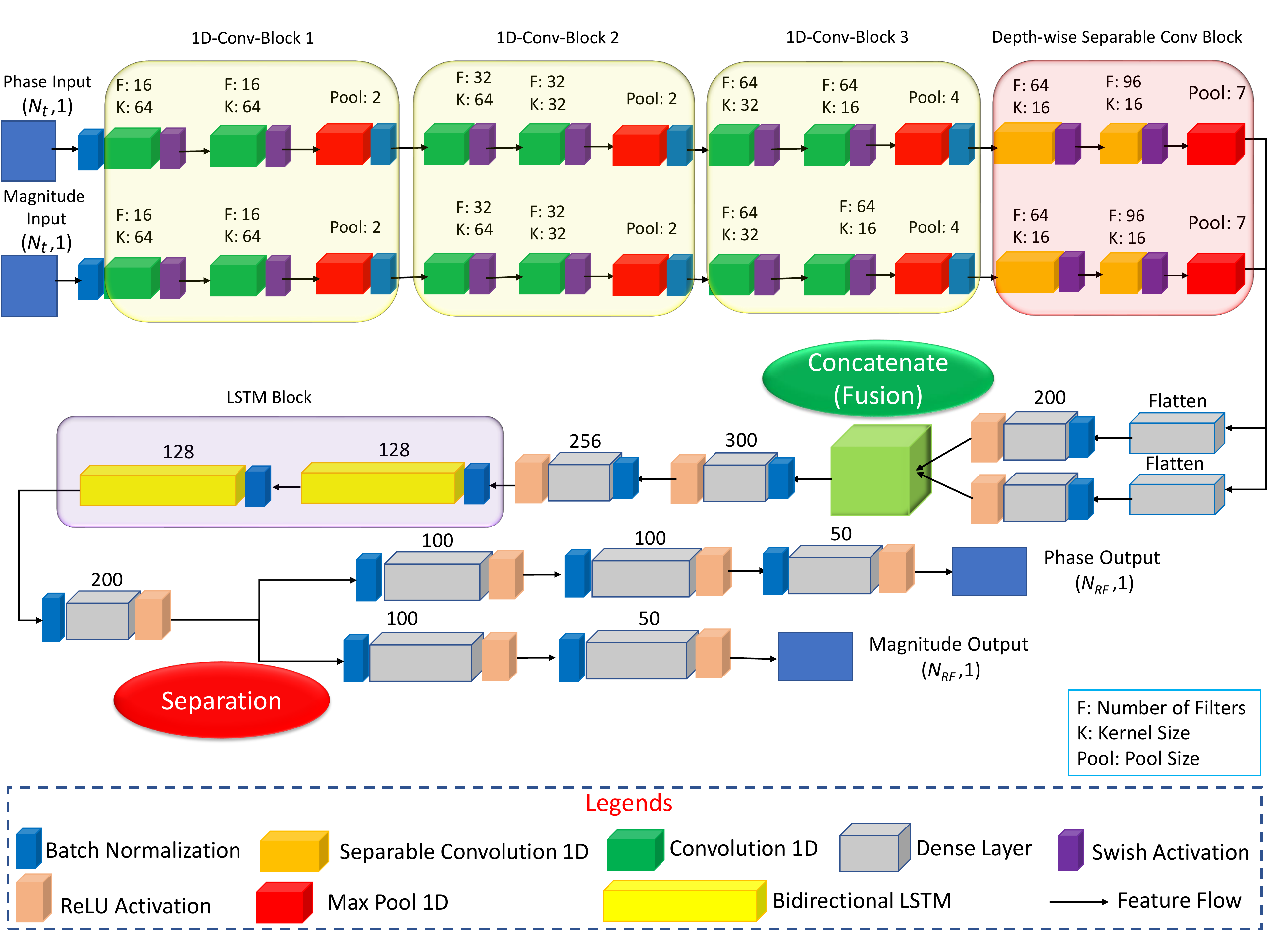}
    \caption{Architecture of the proposed 1D CNN-LSTM-based fusion separation DNN.}
    \vspace{-2mm}
    \label{model-architecture}
\end{figure*}
The primary goal of our neural network-based approach is to minimize latency as well as achieve a high sum-rate. With that in mind, we propose a lightweight 1D CNN-LSTM-based architecture with a comparatively smaller number of parameters than other contemporary networks and estimate the beamforming matrices with high accuracy. A vivid schematic of our proposed network is shown in Figure 5. Here, the dropout layers are omitted for visual clarity. The input to the neural network is two vectors of size (\textit{N\textsubscript{t}}, 1) \footnote{At the receiver end, the input would be (\textit{N\textsubscript{r}}, 1).}. The normalized resolved components of \textbf{F}\textsubscript{opt} columns constitute the inputs. The phase and magnitude inputs are separately passed onto two independent input layers. In each branch, after the input layer, we have three blocks of 1D convolution, where each block has two convolution layers followed by a max-pooling layer. The first convolutional block does not have zero-padding to avoid superfluous noise insertion in the initial convolutional layers. Nevertheless, the rest of the convolutional blocks apply zero-padding to prevent the feature dimensions from becoming too small. After performing thousands of instances of model training, it is observed that using fewer than three blocks reduces the accuracy of the DNN. Conversely, using four or more blocks generates an unnecessary computational burden and gradient underflow with no noticeable improvement in accuracy. Then we use a 1D depth-wise separable convolution block, which again consists of two depth-wise separable convolution layers followed by a pooling layer. We perform extensive trials and errors to determine the specific number of separable convolutional layers used in the model.

\par After extracting features from the convolutions, we flatten the branches and add a dense layer. We then concatenate (fusion) both branches and pass them through two more dense layers. These dense layers change the dimension of the output features from the convolutional network and appropriately shape the inputs to the subsequent LSTM layers. We then use a bidirectional LSTM block with two LTSM layers. Empirically, after performing thousands of training simulations, we recognize that only two LSTM layers are sufficient to provide optimal accuracy, as observed in Figure \ref{LSTM}. Since LSTM layers are computationally pretty expensive , using as few as possible is desirable. Subsequently, the two branches are again separated, and each branch is processed through a block of dense layers to generate the final outputs. The final outputs are two vectors of size (\textit{N}\textsubscript{RF}, 1), one for the magnitude and one for the phase \footnote{Here, \textit{N}\textsubscript{RF} would be $N^t_{RF}$ and $N^r_{RF}$ at the transmitter and receiver end, respectively.}. 
\par The outputs of each convolution layer are passed through a Swish activation function, which performs slightly better than the ReLU for Convolution \cite{article}. The swish activation function for any input $\xi$ is given by

\begin{equation}
    swish(\xi )=\xi  \times \frac{1}{1+e^{-\beta \xi }},
\end{equation}
where $\beta$ is a trainable parameter.
\par Additionally, we use ReLU activation after each dense layer. After every convolution block, we use a 1D spatial dropout layer with a dropout value set to 0.03, while after every dense block, we use Alpha-dropout with a dropout value of 0.04. These dropout layers ensure better generalization and avoid overfitting \cite{JMLR:v15:srivastava14a}. In addition, batch-normalization layers are used after each block and L1-L2 regularizers are incorporated into each layer to accelerate the training process \cite{regularizer}. It can be observed that the separation network for the phase has more depth. This is because, empirically, we observe that SE is relatively more sensitive to slight changes in phase than in magnitude after much trial and error. Therefore, estimating the phase more precisely using a deeper network is vital. We also formulate two other lighter network architectures, with all the layers being the same as described above, the only difference being the number of neurons in the layers. Hence, the three architectures have a different number of total parameters. The largest one has 2.4 million parameters, which we call DNN-Large. The other networks with 1.3 million and 864,000 parameters are aptly named DNN-Medium and DNN-Small, respectively. The number of neurons, kernel sizes, pool sizes and the number of filters in the layers of each model are determined through extensive simulations. 

\subsection{Dataflow in the Network}
\begin{figure}[t]
    \includegraphics[width=\linewidth]{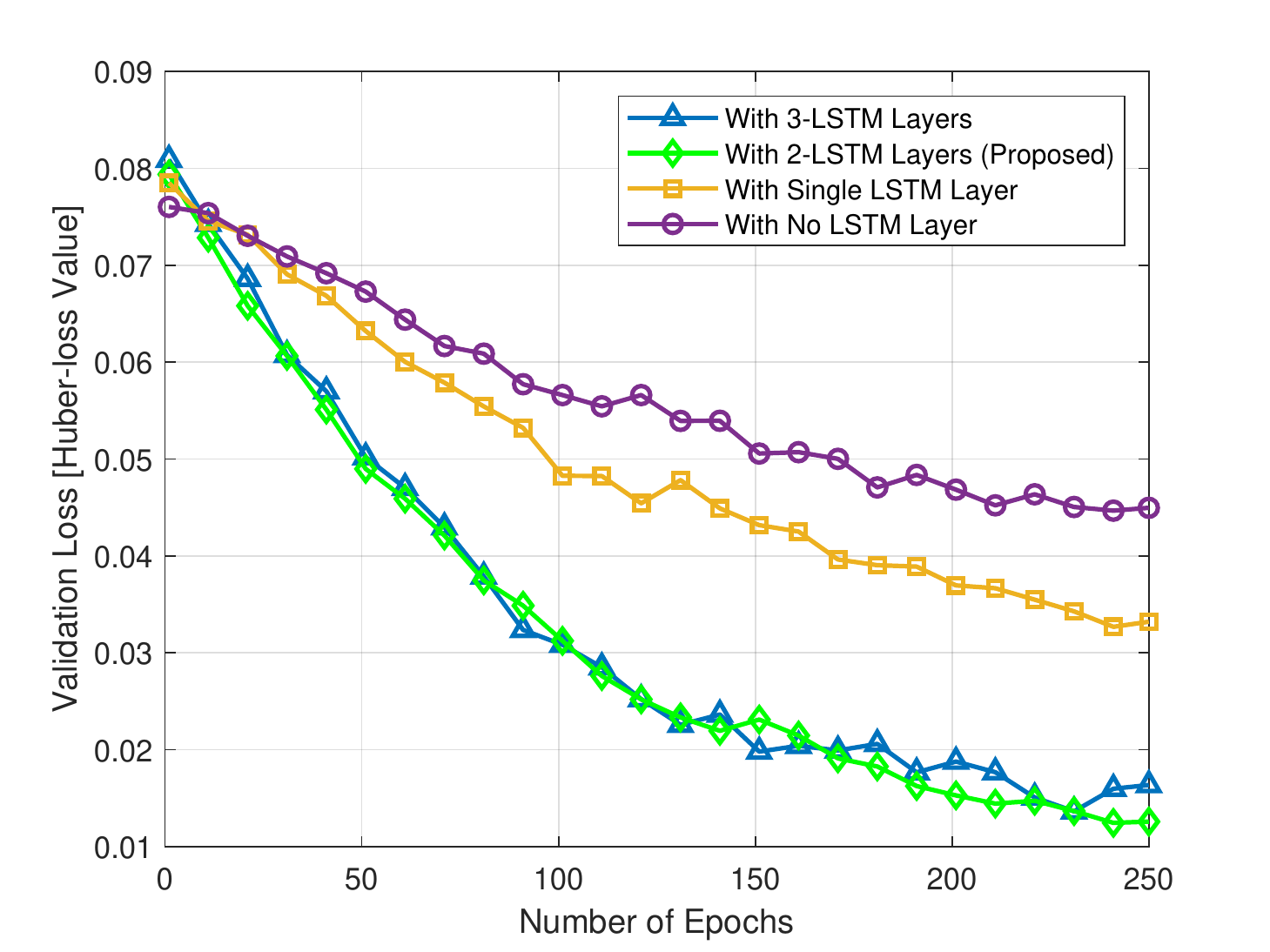}
    \caption{Ablation study for the LSTM layers.}
    \vspace{-2mm}
    \label{LSTM}
\end{figure}

\begin{figure*}[ht]
    \includegraphics[scale=0.5,width=\linewidth]{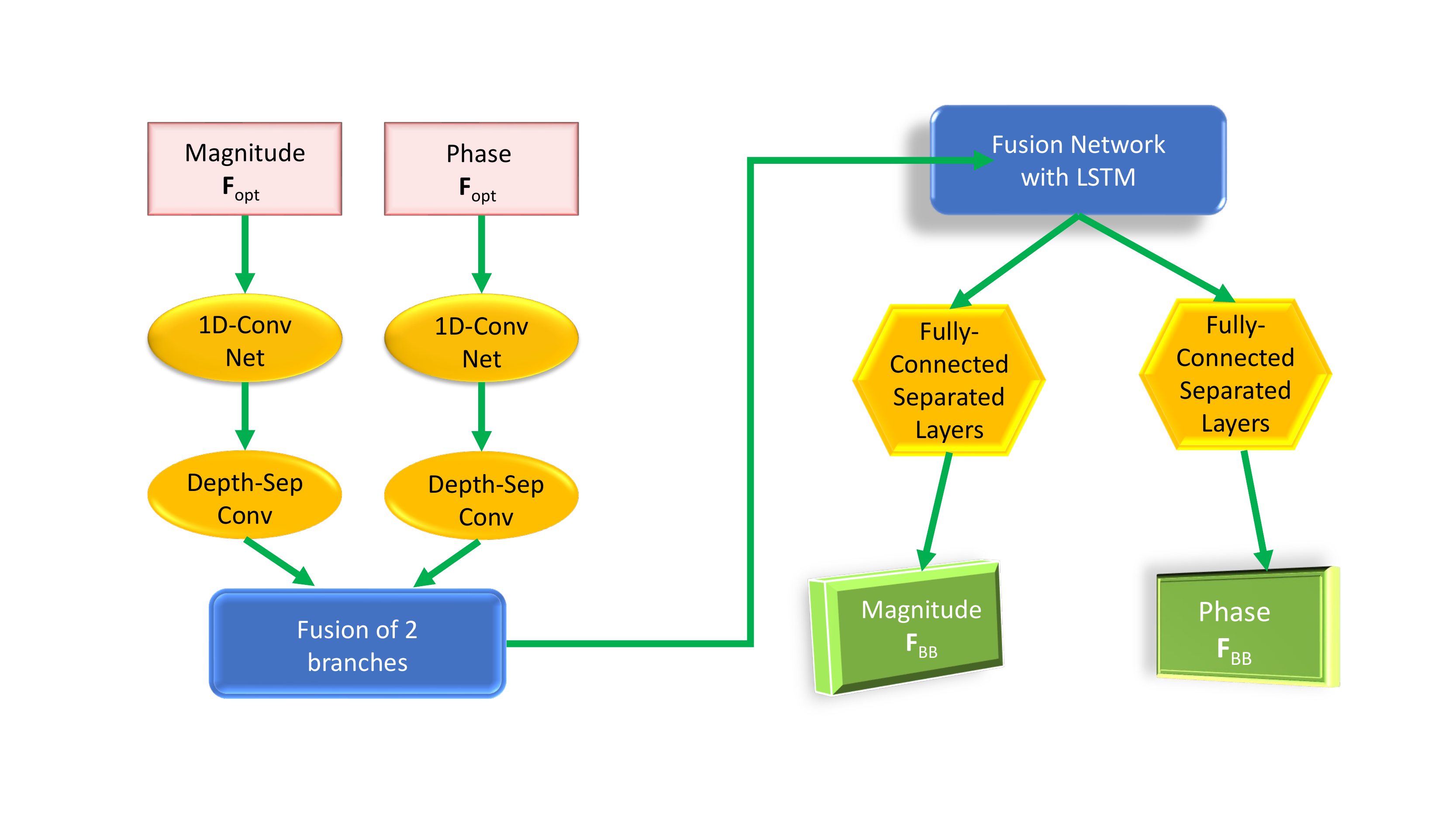}
    \caption{Data-flow in the proposed 1D CNN-LSTM-based architecture.}
    \vspace{-2mm}
    \label{dataflow}
\end{figure*}
In order to efficiently and precisely compute the beamforming matrices \textbf{F}\textsubscript{BB} and \textbf{W}\textsubscript{BB}, we compute their constituent columns separately. We first pass the columns of \textbf{F}\textsubscript{opt} as input to our neural network. The phase and magnitude of elements of \textbf{F}\textsubscript{BB} and \textbf{W}\textsubscript{BB} are independently processed through a 1D-Convolutional network first. This convolutional network extracts unique features from phase and magnitude autonomously. Mathematically, for each feature map output \textbf{o}[i], this can be expressed as
\begin{equation}
    \mathbf{o}[i]= \sum_{n=0}^{N}\\(\boldsymbol{\chi}[i+n]\times \mathbf{k}[n]),
\end{equation}
where $\boldsymbol{\chi}$ is the input data, \textbf{k} is the kernel coefficients, and \textit{N} is the length of the kernel. 
\par Then a depth-wise separable convolutional network is used to simultaneously reduce the total parameters of the network and extract more distinct features from the input data. Since the elements of \textbf{F}\textsubscript{BB} are related to both the phase and the magnitude of \textbf{F}\textsubscript{opt}, the independently extracted features from both the phase and magnitude branches must be combined. This is done in a fusion network consisting of dense and LSTM layers. The LSTM layers are particularly adept at extracting complex relations between sequential and fragmented data. Since the properties of a beamforming matrix depend not only on the values of its constituent elements but also on the position in which the elements are present, we can think of the elements as a sequence of values. The LSTM layers help to extract these very sequential features from the data. The performance of the proposed DNN degrades considerably without the use of LSTM layers, as can be observed from the loss curve in Figure \ref{LSTM}. In this fusion network, features extracted from the phase and magnitude of \textbf{F}\textsubscript{opt} are combined and generate new components, which are now passed on to two separate branches. This separation network is used to predict the phase and magnitude of elements of \textbf{F}\textsubscript{BB} and \textbf{W}\textsubscript{BB} independently. Due to the use of 1D CNN and LSTM along with fusion and separation of two discrete branches as shown in Figure \ref{dataflow}, we name our network as 1D CNN-LSTM fusion-separation network.

\subsection{Post Processing and Generating Beamforming Matrices}
The outputs of the neural networks are normalized vectors which are in fact the phase and magnitude of columns of \textbf{F}\textsubscript{BB} and \textbf{W}\textsubscript{BB}. In order to generate the true baseband matrices, we need to reverse the normalization that is done in the pre-processing. Then the elements of the columns of \textbf{F}\textsubscript{BB} and \textbf{W}\textsubscript{BB} are given by
\begin{equation}
\begin{gathered}
z_P(i)=k_P(i)\times2\pi-\pi \\
z(i)=k_M(i)\cos{z_P(i)}+jk_M(i)\sin{z_P(i)},
\end{gathered}
\end{equation}
where $k_P(i)$ and $k_M(i)$ denote the ith normalized phase and magnitude output from DNN respectively. $z(i)$ represents the ith element of a single column of the digital baseband beamforming matrices. After that, we concatenate the columns together to form the initial matrices \textbf{F}\textsubscript{BBi} and \textbf{W}\textsubscript{BBi}. Now, we find the analog beamforming matrices \textbf{F}\textsubscript{RF} and \textbf{W}\textsubscript{RF} using:
\begin{equation}
\begin{aligned}
  \textbf{F}\textsubscript{RFi} =  \textbf{F}\textsubscript{opt}\dagger \textbf{F}\textsubscript{BBi} \\
  \textbf{W}\textsubscript{RFi} =  \textbf{W}\textsubscript{opt}\dagger \textbf{W}\textsubscript{BBi}.
\end{aligned}
      \end{equation}

However, to ensure the element-wise unit modulus constraints, we apply the following transformation
\begin{equation}
\begin{aligned}
  \textbf{F}\textsubscript{RF} =  \cos{\angle \textbf{F}\textsubscript{RFi}}+ j\sin{\angle \textbf{F}\textsubscript{RFi}}\\
  \textbf{W}\textsubscript{RF} =  \cos{\angle \textbf{W}\textsubscript{RFi}}+ j\sin{\angle \textbf{W}\textsubscript{RFi}}
\end{aligned}.
      \end{equation}

Here, $\angle F\textsubscript{RFi}$ and $\angle W\textsubscript{RFi}$ denote the phase of initial analog beamforming matrices. Again, to satisfy the normalized transmit power constraint of (\ref{4.6}) we utilize the following transformation
\begin{equation}
\begin{aligned}
  \textbf{F}\textsubscript{BB} =  \frac{\sqrt{N_s}\times \textbf{F}\textsubscript{BBi}}{\norm{\textbf{F}\textsubscript{RF}\textbf{F}\textsubscript{BBi}}_F} \\
  \textbf{W}\textsubscript{BB} =  \frac{\sqrt{N_s}\times \textbf{W}_\textsubscript{BBi}}{\norm{\textbf{W}\textsubscript{RF}\textbf{W}\textsubscript{BBi}}_F}
.\end{aligned}
      \end{equation}
      
Thus, we obtain all four beamforming matrices.
\section{Simulation Setup}
\subsection{Simulation Environment}
Google Colaboratory, a Python development environment that runs in the browser utilizing Google Cloud and provides free access to strong graphical processing units (GPU), is used to conduct all the simulations. Our proposed 1D-CNN-LSTM system and accompanying peripherals are implemented in Python 3.7.11 using TensorFlow 2.7.0 and a Tesla T4 GPU supplied by Google Collaboratory. To ensure a true and fair comparison of the execution time of various processes, all codes are executed using the same configuration.
\subsection{Dataset}
We start by generating 100,000 channel realizations (\textbf{H}) and corresponding \textbf{F}\textsubscript{opt}, \textbf{F}\textsubscript{BB}, \textbf{F}\textsubscript{RF}, \textbf{W}\textsubscript{opt}, \textbf{W}\textsubscript{BB} and \textbf{W}\textsubscript{RF}. Then, each \textbf{F}\textsubscript{opt}, \textbf{F}\textsubscript{BB}, \textbf{W}\textsubscript{opt} and \textbf{W}\textsubscript{BB} are resolved into $N_s$ columns. Total number of training samples thus becomes 500,000 for $N_s$ = 5. Generating this huge amount of data using the MO-Alt-Min algorithm would require almost eternity. Hence, most of the data (90\%) is generated using the OMP algorithm and the model is first trained on that data. After that, transfer learning is used to carry the weights from this initial training phase and the model is then trained on MO-Alt-Min generated data, which constitutes only 10\% of the entire dataset. We also generate 10,000 channel realizations to produce the test dataset. Unless otherwise mentioned, all experiments were conducted with $N_t$ = $N_r$ = 256, $N_s$ = 5, $N_{RF}$ = 5, $\lambda$ = 0.1 cm ($f$ = 0.3 THz), and $d$ = 0.05 cm .

\subsection{Simulation Parameters and Model Training}
After data generation is finished, we split the generated training data into a training data set and a validation data set in the typical ratio of 8:2. The training process optimizes the weight of the trainable network parameters through back-propagation, and we choose the Adagrad algorithm as the optimization algorithm. Adagrad, also known as adaptive gradient, permits the learning rate to adapt based on specified parameters. It executes larger changes for infrequent parameters and more minor changes for frequently updated ones. This makes it well-suited for sparse data \cite{duchi2011adaptive}. Additionally, we use Huber loss as our loss function, which is quite well suited to regression tasks as it fuses the best attributes of the mean square error and mean absolute error losses \cite{https://doi.org/10.48550/arxiv.2108.12627}.
For each value $\epsilon$ in $error = y_{true}-y_{pred}$, loss is given by



\begin{equation}
loss = \begin{cases}
          0.5\times \epsilon^2 \quad &\text{if} \, |\epsilon| \le \psi \\
          0.5\times \epsilon^2 + \psi \times (|\epsilon|-\psi) \quad &\text{if} \, |\epsilon| > \psi \\
     \end{cases},
\end{equation}
where `$\psi$' is a constant parameter that determines the exact behavior of the loss towards outliers. We choose a typical value of $\psi$ = 1 for all simulations.
\par
\begin{figure}[t]
    \includegraphics[width=\linewidth]{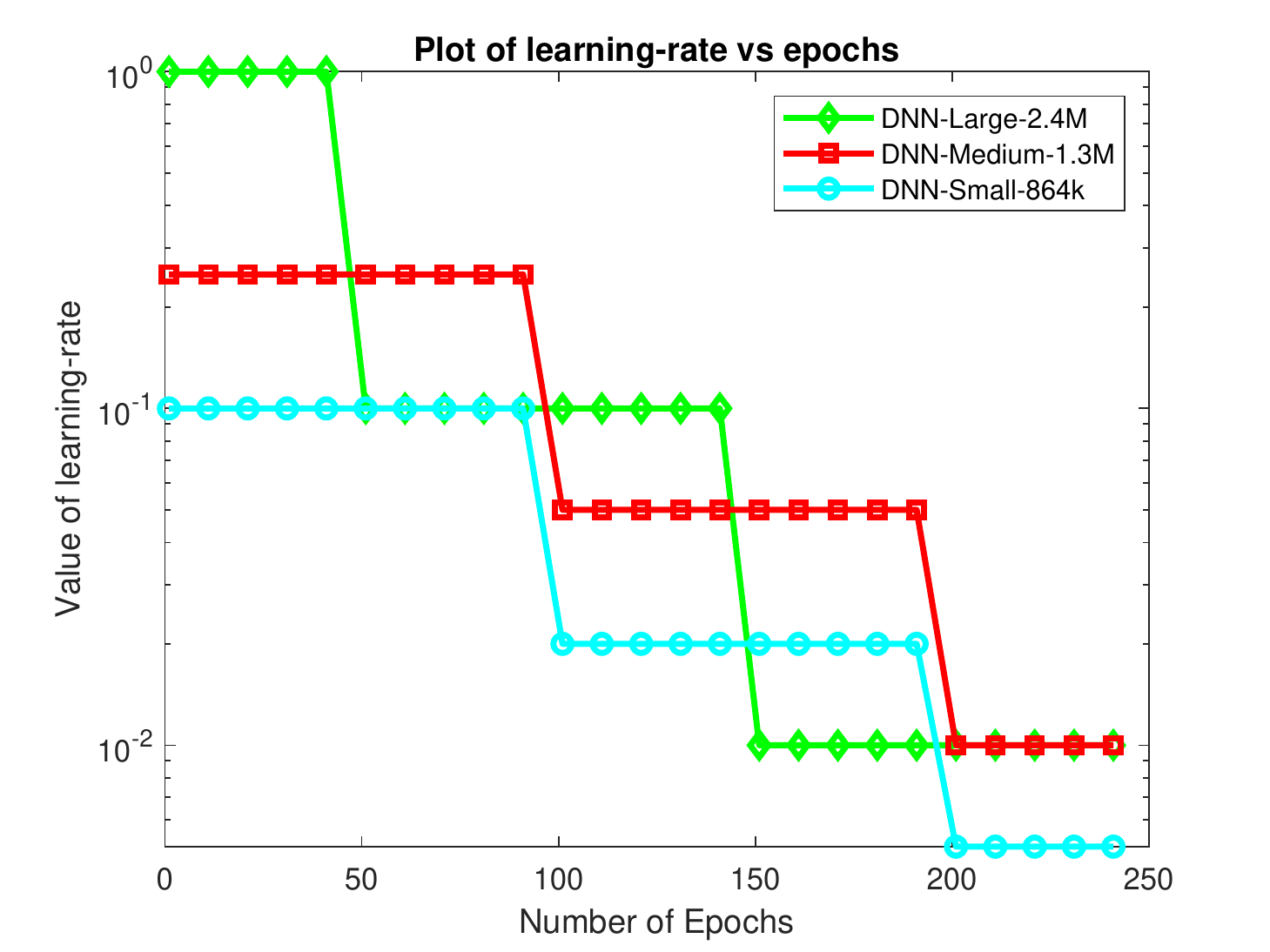}
    \caption{Learning rate used for different model sizes with the number of epochs.}
    \vspace{-2mm}
    \label{learningrate}
\end{figure}

\begin{figure}[t]  \includegraphics[width=\linewidth]{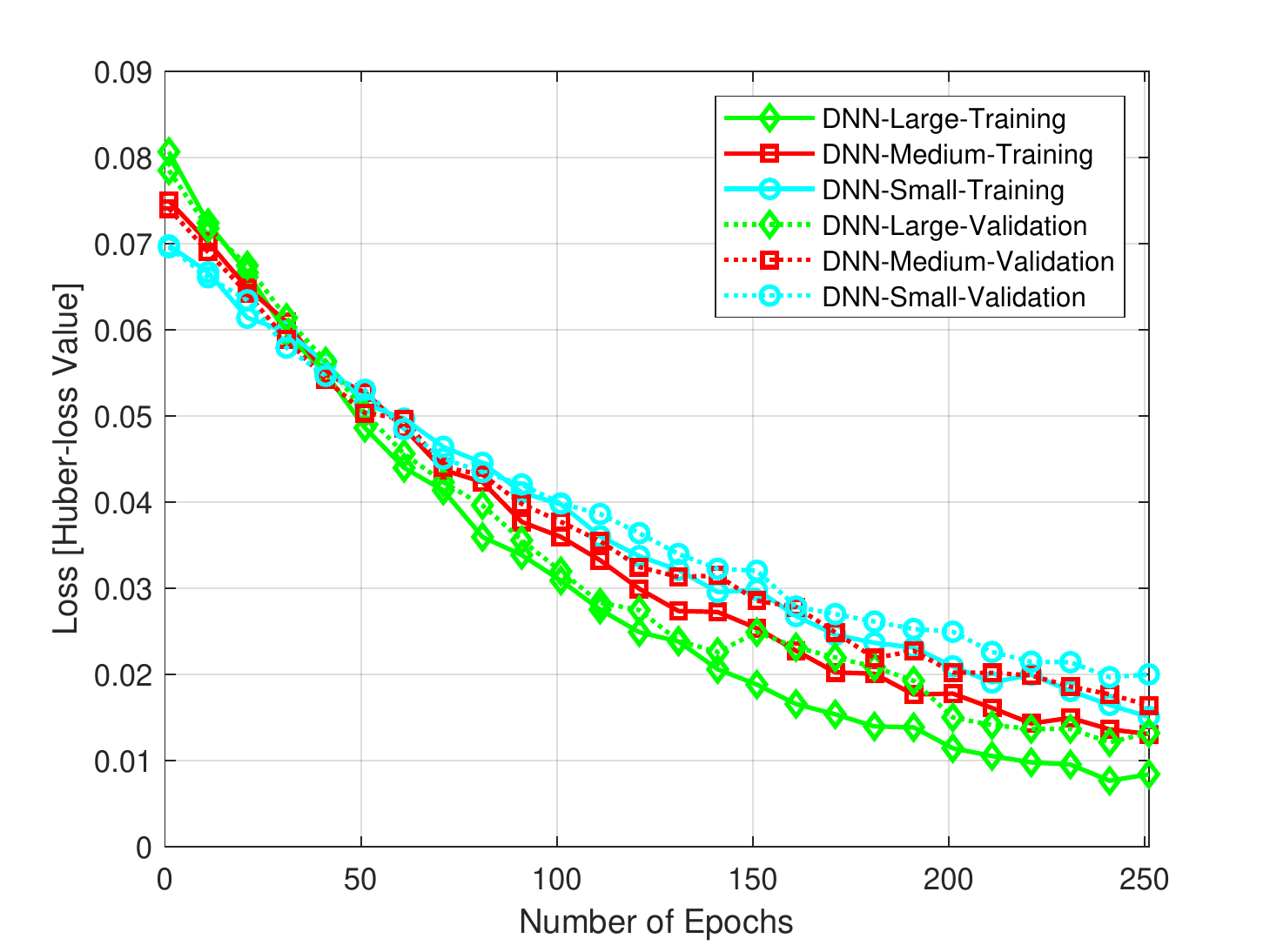}
    \caption{Training and validation loss of different model sizes with the number of epochs.}
    \vspace{-2mm}
    \label{loss_curve}
\end{figure}

 We choose a batch size of 1024, which is the maximum that the employed Tesla T4 GPU supports. The higher the batch size, the better the model predictions. The learning rate, however, is varied throughout the whole training process, and we use a stepwise constant decay scheduler. Figure \ref{learningrate} shows how the proposed scheduler adapts the learning rate based on the training progress. Scheduling the learning allows the model to reach a stable minima much faster than a constant learning rate. As we can observe from figure \ref{learningrate}, the learning rate also varies according to the size of the model. The largest model, with 2.4 million parameters, has a higher learning rate because it starts at a higher point in the loss curve than the smaller models as shown in Figure \ref{loss_curve}. Although our network has dropout layers, we also apply both L1 and L2 penalties for further regularization \cite{regularizer}. Here, the values of L1 and L2 are empirically chosen to be 0.000001. We monitor the learning progress of the network through appropriate validation data set performance metrics. The performance metrics for the validation set are the mean square error and the mean absolute percentage error.

\section{Results}
\begin{figure*}[ht]
\centering
    \includegraphics[width=\linewidth]{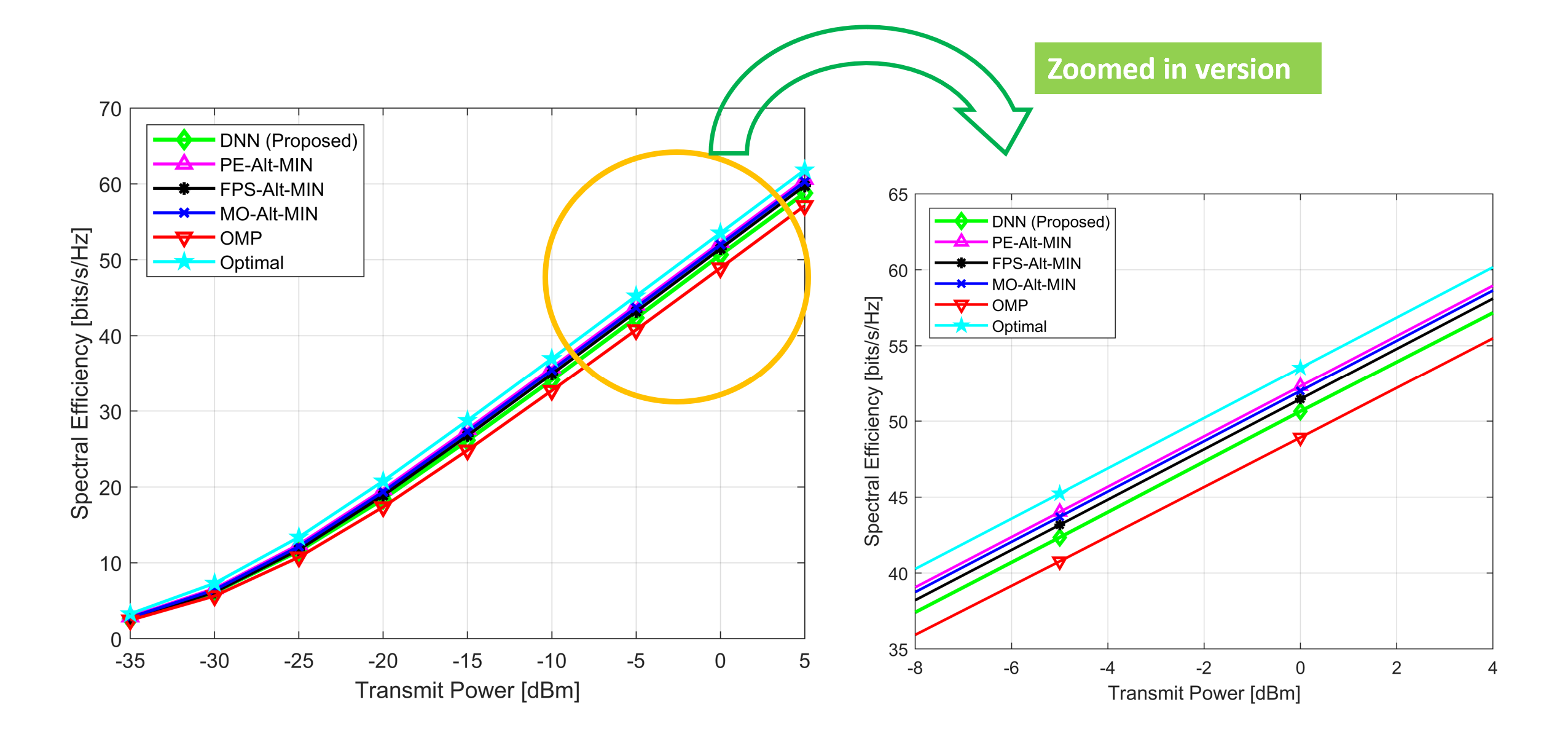}
    \caption{SE of different algorithms for varying transmit power with \textit{N\textsubscript{t}} = \textit{N\textsubscript{r}} = 256 and \textit{N\textsubscript{s}} = \textit{N\textsubscript{RF}} = 5.}
    \vspace{-2mm}
    \label{SE_vs_TxP}
\end{figure*}
In this section, we present the simulation results and corresponding analysis to validate the proposed beamforming system. We evaluate the performance of our proposed system in terms of SE and computational time. Moreover, we compare our approach to the already established beamforming algorithms and analyze their advantages and drawbacks.
\subsection{Spectral Efficiency Comparison}
In this segment, we compare the SE performance of our proposed system with the existing beamforming algorithms such as the MO-Alt-Min, OMP, PE-Alt-Min etc.. We also show the SE achieved by fully digital beamforming for benchmarking purpose. 

In order to find the SE, we first calculate the beamforming matrices using the existing beamforming algorithms such as the MO-Alt-Min, OMP, PE-Alt-Min, and FPS-Alt-Min for all the 10,000 test channel realizations. Additionally, the fully digital optimal beamforming matrices are found simply by the singular value decomposition of the channel matrices. Subsequently, the beamforming matrices for our proposed method are generated using the approach described in section IV. Now, the SE of all the algorithms is computed using their respective matrices. We plot the SE along with the transmit power to show the potency of our proposed scheme, which is presented in Figure \ref{SE_vs_TxP}. It can be seen that the proposed method achieves almost the same SE as that of the Alt-Min algorithms and outperforms the OMP algorithm by quite a margin. To observe the slight differences in achieved SE, we take a closer look at the plot and see that our neural network based beamformer approximately achieves only 1.6 bits/s/Hz less than the most spectrally efficient MO-Alt-Min and PE-Alt-Min. Unsurprisingly, we also observe that as the transmit power increases, the SE of all the algorithms increases, but they are always less than the SE achieved by the fully digital beamformer\footnote{We designate the fully digital beamformer as 'Optimal' in all relevant figures.}. The fully digital beamformer requires as many RF chains as the number of antennas it has. This is both costly in terms of hardware and also has much higher power consumption. Through hybrid beamforming, we are effectively trading off a small amount of SE for a large reduction in power consumption and hardware costs.
\subsection{Relative Computational Time}
Our primary objective is to reduce the computation time required for real-time hybrid beamforming in order to support THz communication in 6G networks. In this sub-section, we compare the performance of our proposed DNN-based beamforming approach to that of the well-known Alt-Min algorithms in terms of computational time.
\begin{figure}[t]
    \includegraphics[width=\linewidth]{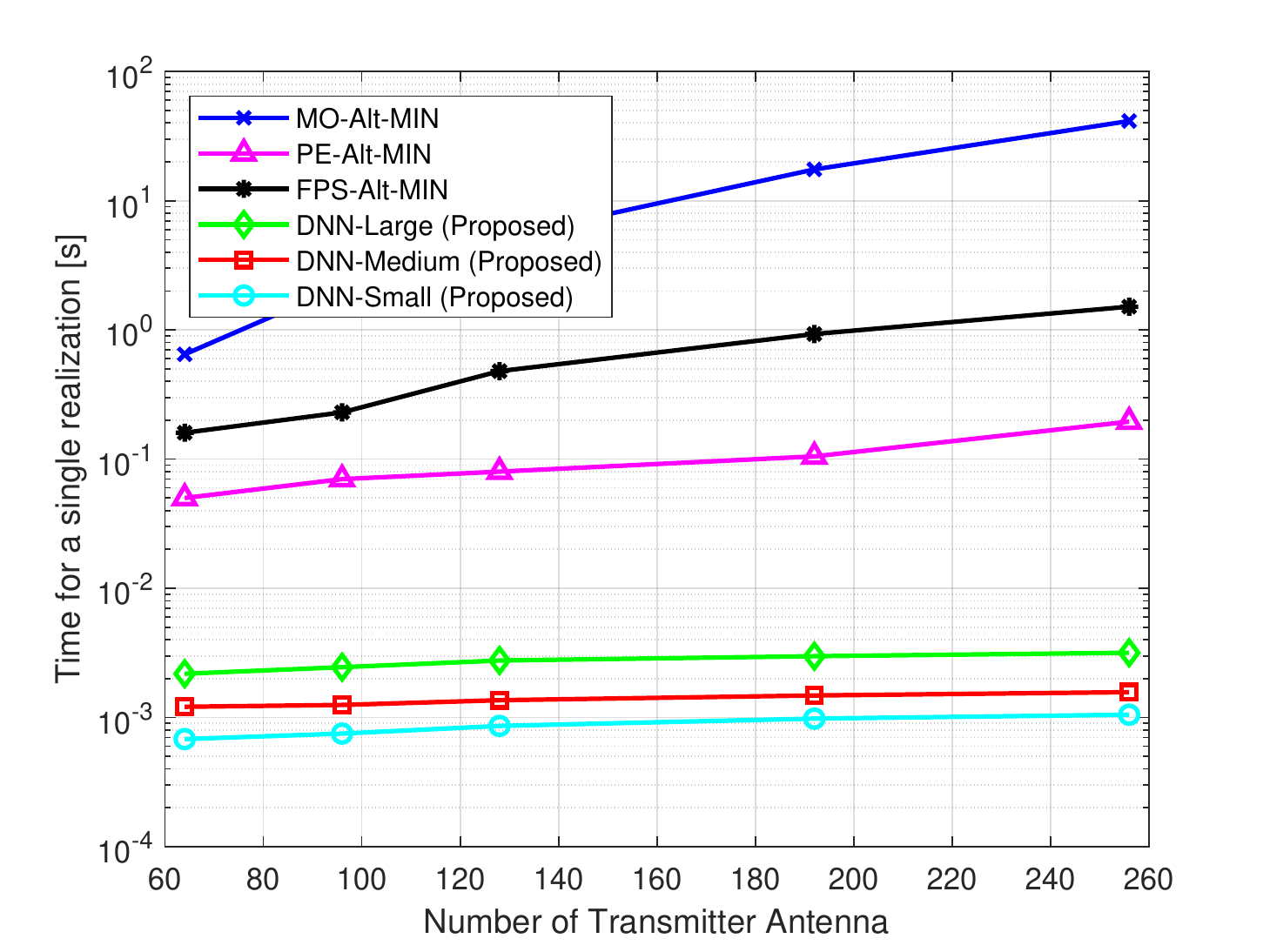}
    \caption{Computational time required by different beamforming algorithms with the number of transmitter antennas. The number of receiver antenna is always kept the same as the number tranmitter antenna (\textit{N\textsubscript{t}} = \textit{N\textsubscript{r}}).}
    \vspace{-2mm}
    \label{Tx_vs_time}
\end{figure}
In order to obtain infallible validity, the computational time for each algorithm is determined by simulating 10,000 channel realizations and then averaging the results. The results are presented in Figure \ref{Tx_vs_time}. We observe that our proposed DNN-based beamformer is almost 100 times faster than the fastest Alt-Min algorithm, the PE-Alt-Min. As the number of transmitter antennas increases, the computational requirements of the Alt-Min algorithms rise quite steeply because these algorithms rely on numerical iterations, which become more computationally expensive as the number of input variables involved increases. However, for our proposed approach, increasing the number of antennas only requires the input layer to be changed, while the rest of the network architecture remains exactly as predefined. Hence, only a slight increase in inference time is observed. This is a major advantage of the proposed neural network-based beamforming scheme. We also observe that the computational time of the DNN-Small network is the lowest, which is due to the fact that it has the least number of parameters. However, we shall see in the subsequent discussion that this reduction in computational cost comes at the price of lower SE. For a more comprehensive comparison between the three neural network architectures, we examine their relative time-gain with respect to the MO-Alt-Min algorithm. Figure \ref{Relative_Time} reveals that for 256 transmitter antennas, the largest DNN model is approximately 10,000 times faster than the MO-Alt-Min and the smallest one is around 40,000 times faster.
\begin{figure}[t]
    \includegraphics[width=\linewidth]{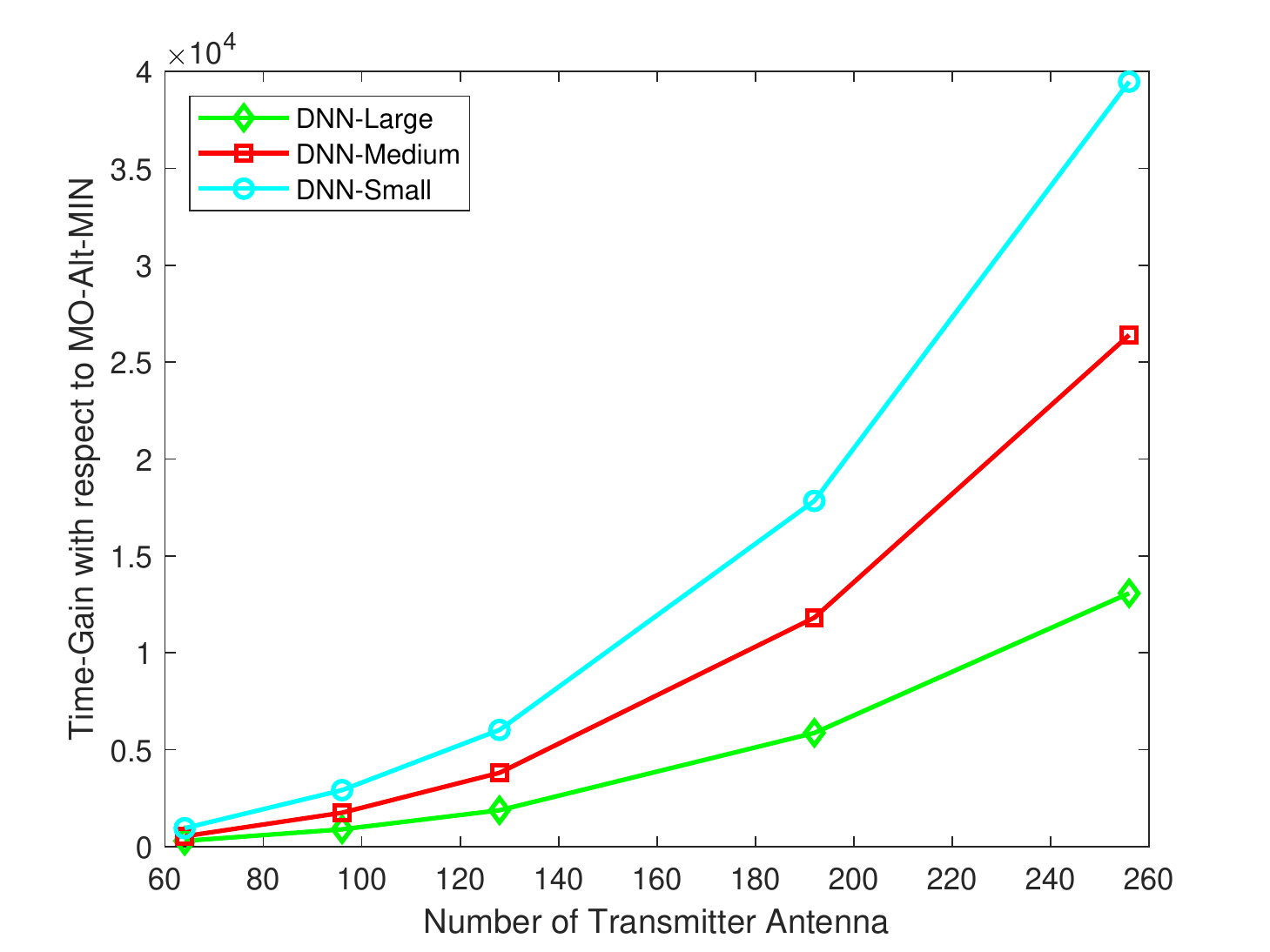}
    \caption{Relative computation gain of different model sizes with the number of transmitter antenna (Here, \textit{N\textsubscript{r}} = \textit{N\textsubscript{t}}).}
    \vspace{-2mm}
    \label{Relative_Time}
\end{figure}
\subsection{Performance with varying numbers of RF chains}
\begin{figure}[t]
    \includegraphics[width=\linewidth]{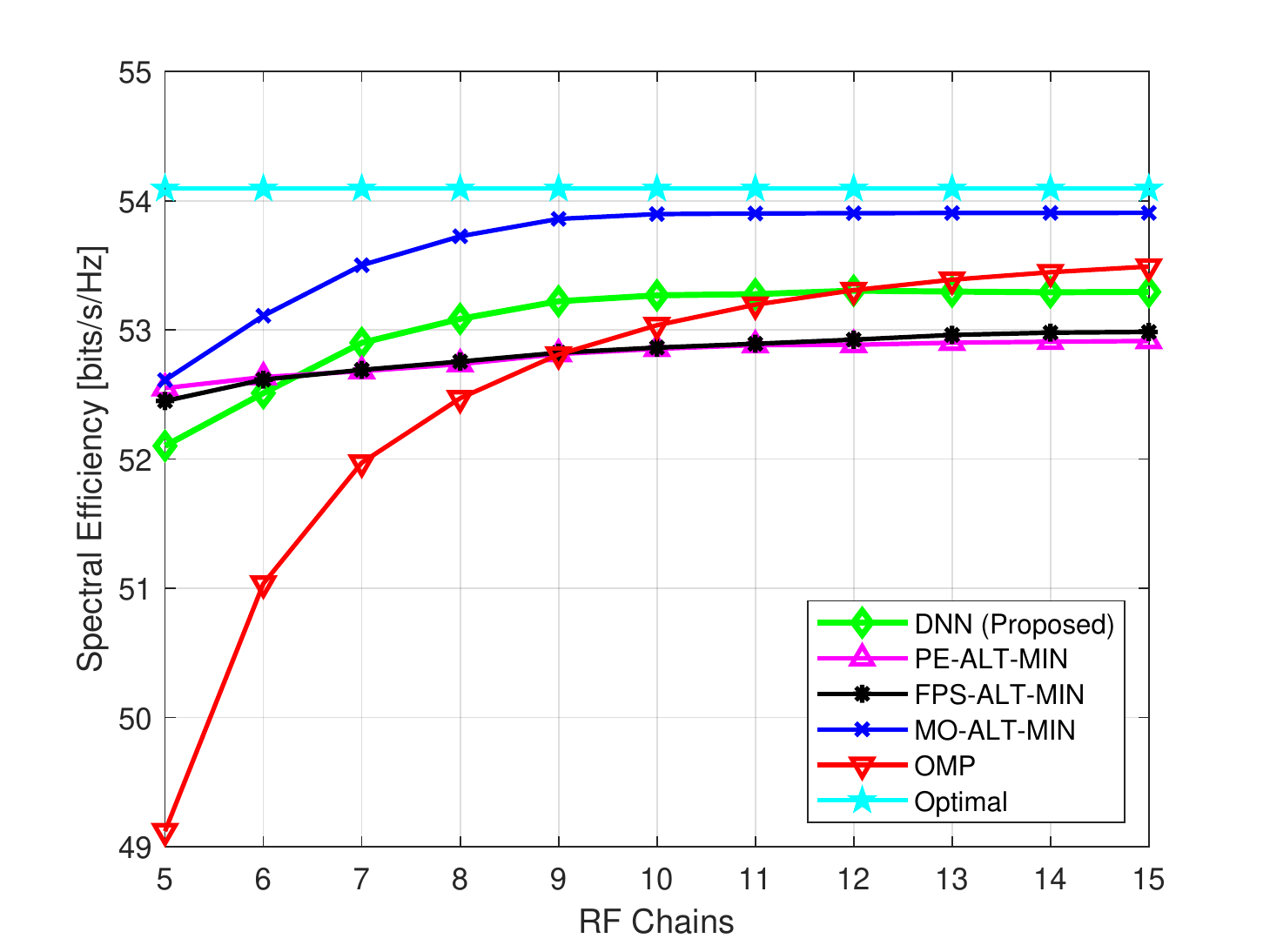}
    \caption{SE of different algorithms with respect to the number of RF chains.}
    \vspace{-2mm}
    \label{SE_vs_RF}
\end{figure}
One of the vital considerations in hybrid beamforming design is the number of RF chains at the transmitter and the receiver. Reducing the number of RF chains as compared to fully digital beamforming has been the primary reason hybrid beamforming research has blossomed. With that in mind, one of the major deliberations in developing our proposed approach is to maintain high SE even when the number of RF chains available varies. Here, we demonstrate the results achieved by our proposed scheme even when the number of RF chains varies substantially. Simulation results in Figure \ref{SE_vs_RF} show that our DNN model outperforms all other algorithms except the MO-Alt-Min in most scenarios. Here, the transmit power is kept constant at 0 dBm and the number of RF chains is varied from 5 to 15. Here, we have kept the minimum number of RF chains at 5 since $N_{RF}< N_s$ is not possible. As the data for the DNN training is generated using the MO-Alt-Min, it is not possible for our supervised learning method to outperform its own data-generating algorithm. Nevertheless, our proposed neural network achieves almost the same SE as that of the MO-Alt-Min even when the number of RF chains varies significantly. It has been proven that when the number of RF chains reaches twice the number of data streams, the SE of hybrid beamforming saturates and becomes almost the same as that of fully digital beamforming \cite{7389996}. In our simulations, there are five independent data streams. Unsurprisingly, we observe that for 10 or more RF chains, the sum-rate of all the algorithms saturates and is almost similar to that of the optimal fully digital beamformer. As the training data of our model comes from MO-Alt-Min, it also achieves an unfluctuating SE when the number of RF chains is increased beyond 10.
\subsection{Trade-off between spectral efficiency and model-size}
\begin{figure}[t]
    \includegraphics[width=\linewidth]{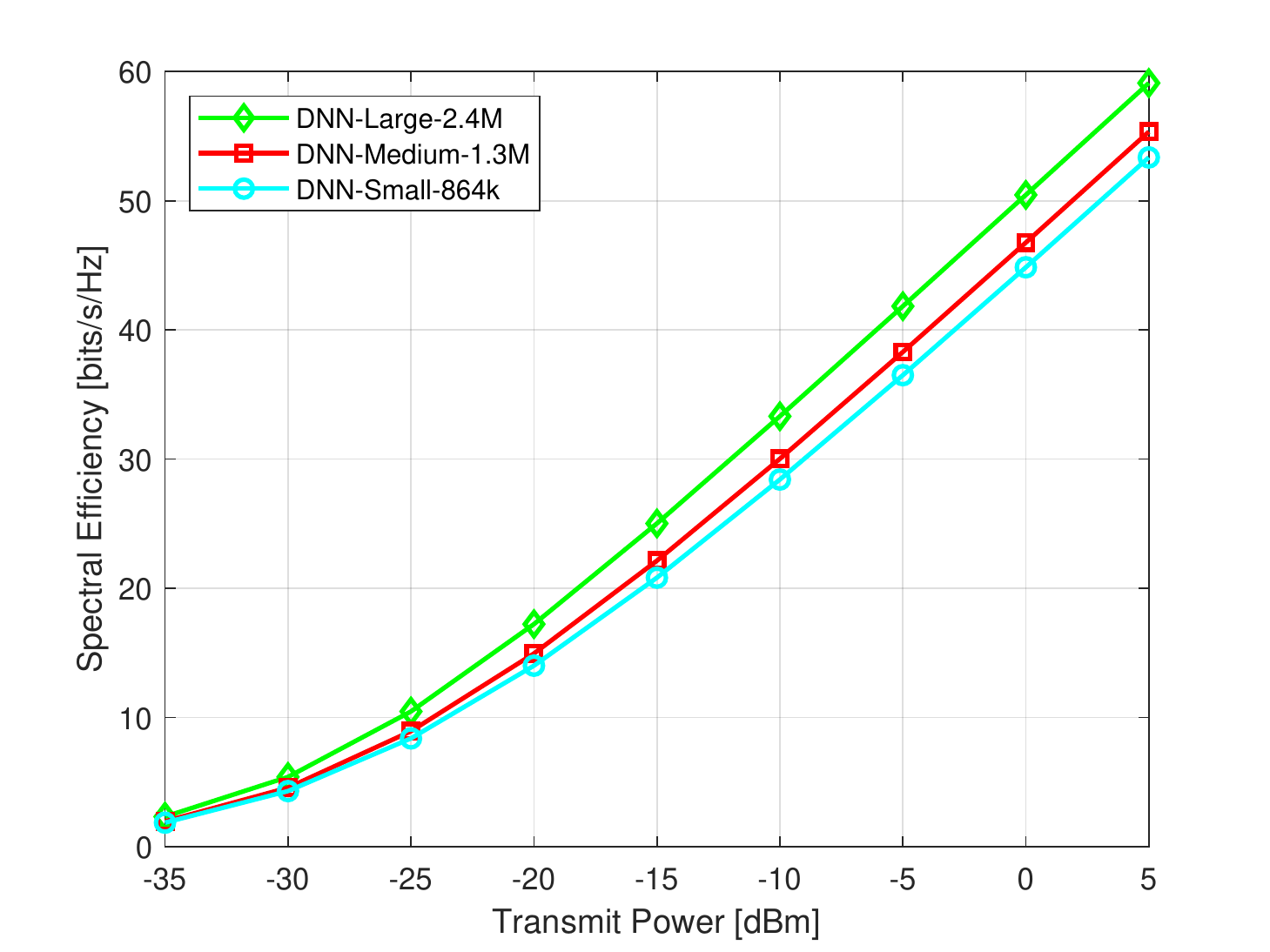}
    \caption{SE of different model sizes for varying transmit power with \textit{N\textsubscript{t}} = \textit{N\textsubscript{r}} = 256 and \textit{N\textsubscript{s}} = \textit{N\textsubscript{RF}} = 5.}
    \vspace{-2mm}
    \label{Model_size}
\end{figure}
As described briefly in the earlier discourse, we propose three slightly different neural networks with a distinct number of total parameters to explore the potential trade-off that arises between the achieved SE and the computational gain. As shown in Figure \ref{Relative_Time}, the smallest DNN model (DNN-Small) is almost four times as fast as the largest one (DNN-Large) due to its significantly fewer number of parameters, which generates a much lower computational overhead. However, this benefit comes at the expense of reduced SE, which becomes evident from Figure \ref{Model_size}. The number of parameters of the model is decreased simply by reducing the number of neurons in each layer. Hence, the smaller models rely on fewer features to make predictions about the beamforming matrices. This makes the predictions less accurate than those of the largest one and consequently results in a lower SE. This trade-off can be well utilized when there are different QoS requirements and in various distinct network scenarios. For instance, if a high data rate is required with no stringent constraints on latency, then the largest model can be deployed to enhance the SE. On the other hand, if latency constraints are more rigorous, such as in ultra-reliable low-latency communications (URLLC), we may opt for the lightest DNN-Small model, which has only 864,000 parameters.
\section{Conclusion and Future Work}
This paper has developed a novel 1D-CNN-LSTM-based fusion-separation DNN for fully connected hybrid beamforming, and contextual mathematical modelling has also been shown. Specifically, we have devised a beamforming method to enable UM-MIMO communications at THz frequencies by employing some of the most influential and cutting-edge deep learning techniques. To validate our claim, we have undertaken extensive simulations to illustrate the efficacy and performance of our approach. We have analyzed the numerical results in terms of SE and computational complexity by varying pertinent parameters such as the number of antennas, the number of RF chains, and the transmit power. The simulation results have proven that the proposed approach can achieve almost the same SE as achieved by the well-known MO-based beamforming algorithms while significantly outperforming them in computational gain. In the future, we will pursue research into partially connected beamforming systems, such as dynamic arrays of sub-arrays, considering imperfect CSI and energy efficiency.


\normalem
\bibliographystyle{IEEEtran}
\bibliography{bibliography.bib}

\begin{IEEEbiography}[{\includegraphics[width=1in,height=1.25in,clip,keepaspectratio]{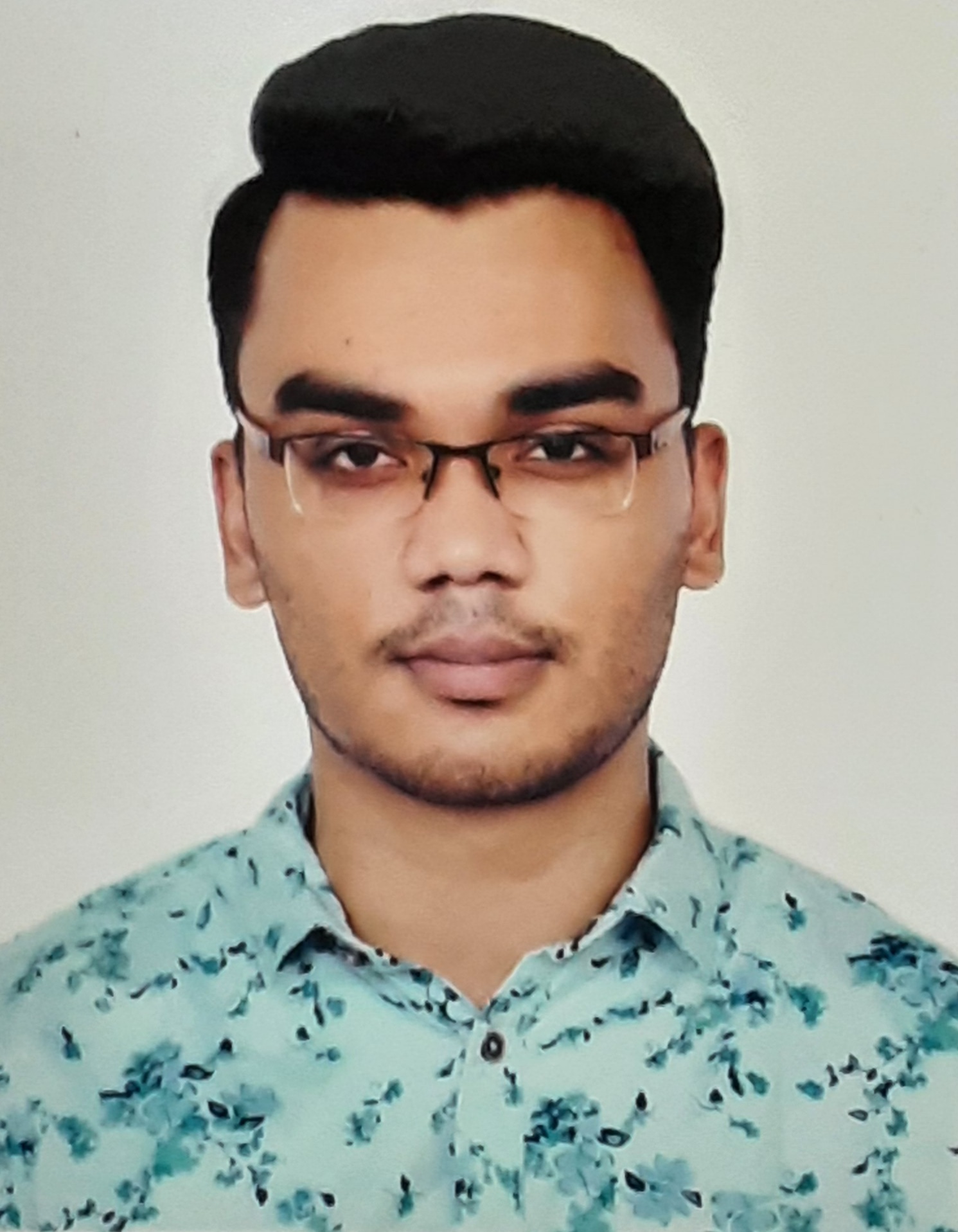}}]{Rafid Umayer Murshed } was born in Sylhet, Bangladesh. He received his BSc degree in Electrical and Electronics Engineering (EEE) from Bangladesh University of Engineering and Technology (BUET), Bangladesh in May 2022. Currently, he is working as a research assistant at BUET-Japan Institute of Disaster Prevention and Urban Safety (JIDPUS), BUET where he is working on developing a disaster-resilient delay-sensitive telecommunication network for earthquake early warning broadcast. His area of interest includes applications of deep-learning and reinforcement learning in wireless communication technologies such as UM-MIMO, beam-forming, resource allocation, reconfigurable intelligent surfaces, etc., as well as in signal processing and Internet-of-Things.
\end{IEEEbiography}

\begin{IEEEbiography}[{\includegraphics[width=1in,height=1.25in,clip,keepaspectratio]{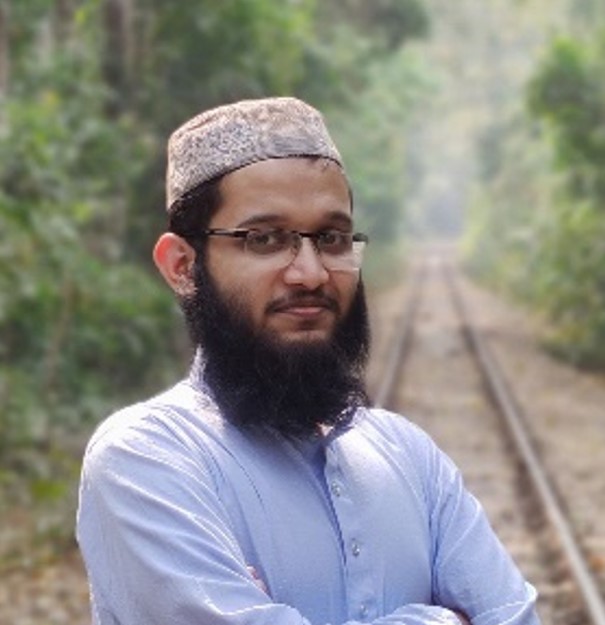}}]{Zulqarnain Bin Ashraf} was born in Dhaka, Bangladesh in October 1996. He received his BSc degree in Electrical and Electronic Engineering from Bangladesh University of Engineering and Technology (BUET), Bangladesh, in May 2022. His research interests include 5G and 6G wireless technologies such as Beamforming, Terahertz Communications, Massive MIMO and Deep Learning.
\end{IEEEbiography}

\begin{IEEEbiography}[{\includegraphics[width=1in,height=1.25in,clip,keepaspectratio]{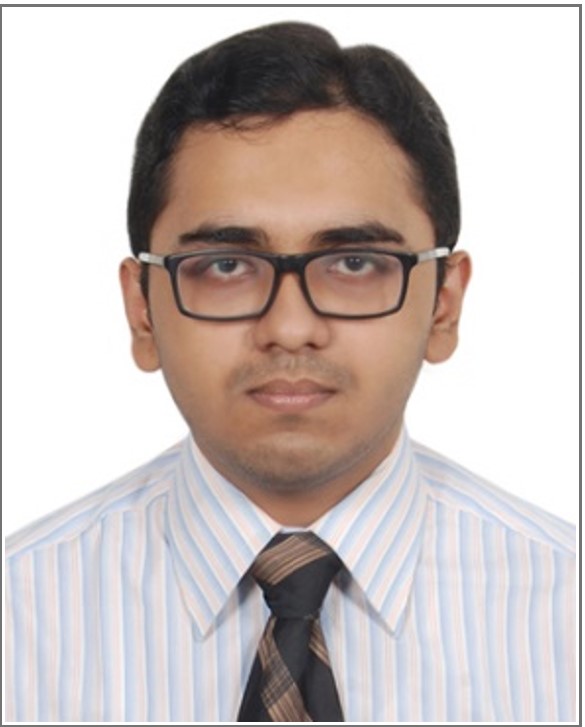}}] {Abu Horaira Hridhon} graduated in 2022 with a B.Sc. in Electrical and Electronic Engineering (EEE) from Bangladesh University of Engineering and Technology (BUET), Bangladesh. He is currently enrolled in the same university's M.Sc. program in Communication and Signal Processing. His current research interests are in wireless communication, computer vision, signal processing, and federated learning.
\end{IEEEbiography}

\begin{IEEEbiography}[{\includegraphics[width=1in,height=1.25in,clip,keepaspectratio]{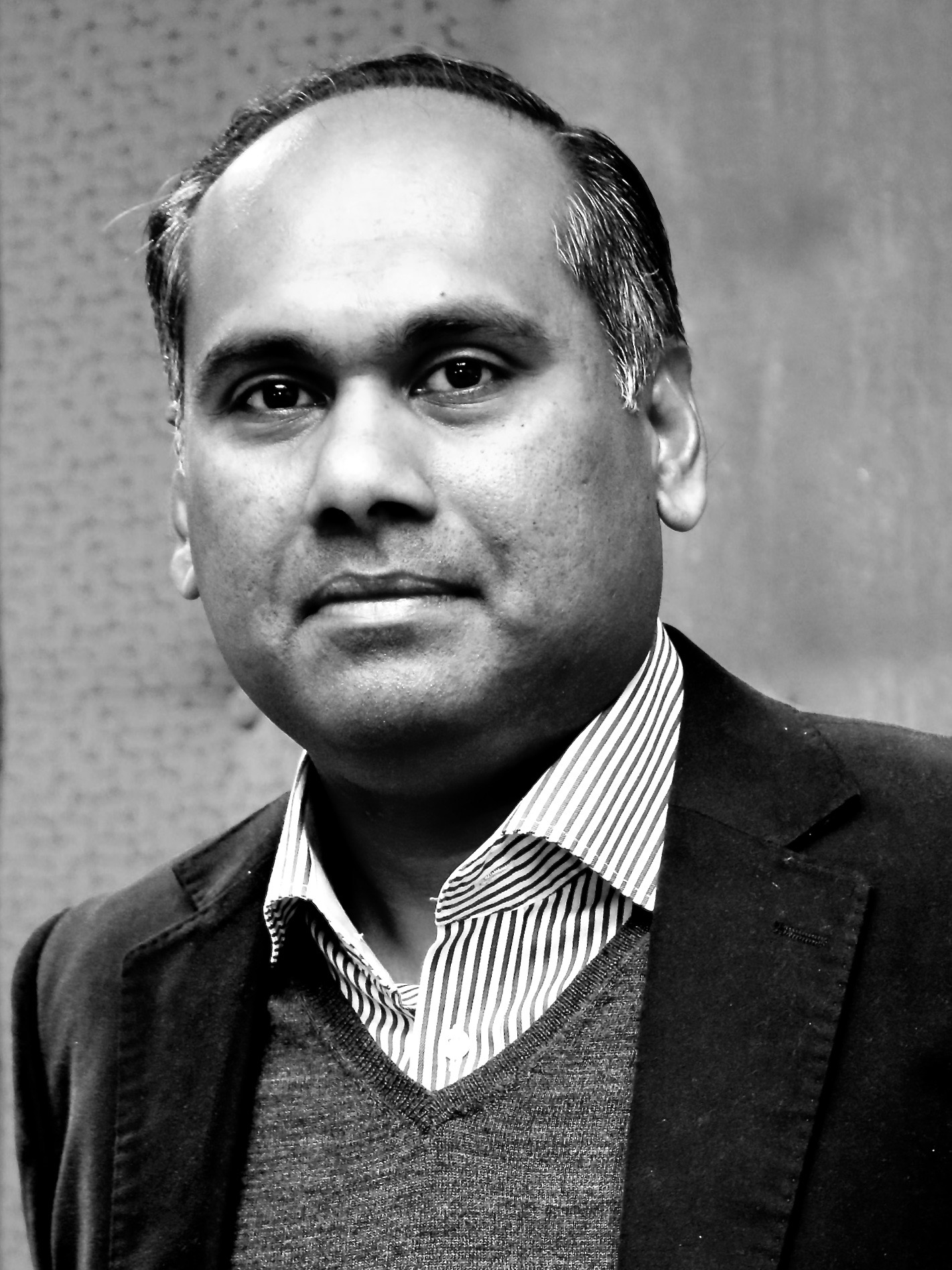}}]{Kumudu Munasinghe} (S’03–M’08) holds a PhD degree in Telecommunications Engineering from the University of Sydney. He is currently an Associate Professor in Network Engineering, leader of the IoT Research Group at the Human Centred Research Centre, University of Canberra. His research focuses on Next Generation Mobile and Wireless Networks, Internet-of-Things, Green Communication, Smart Grid Communications, and Cyber-Physical-Security. Prof. Munasinghe has over 100 refereed publications with over 1250 citations (h-index: 20) in highly prestigious journals, conference proceedings, and two books to his credit. He has secured over \$1.7 million dollars in competitive research funding by winning grants from the Australian Research Council (ARC), the Commonwealth and State governments, the Department of Defense, and industry. He has also won the highly prestigious ARC Australian Postdoctoral Fellowship, served as a co-chair for many international conferences, and served as an editorial board member for a number of journals. His research has been highly commended through many research awards, including two VC's Research Awards and three IEEE Best Paper Awards. He is currently a member of the IEEE, a charted professional engineer, an engineering executive, and a companion (fellow status) of Engineers Australia.
\end{IEEEbiography}

\begin{IEEEbiography}[{\includegraphics[width=1in,height=1.25in,clip,keepaspectratio]{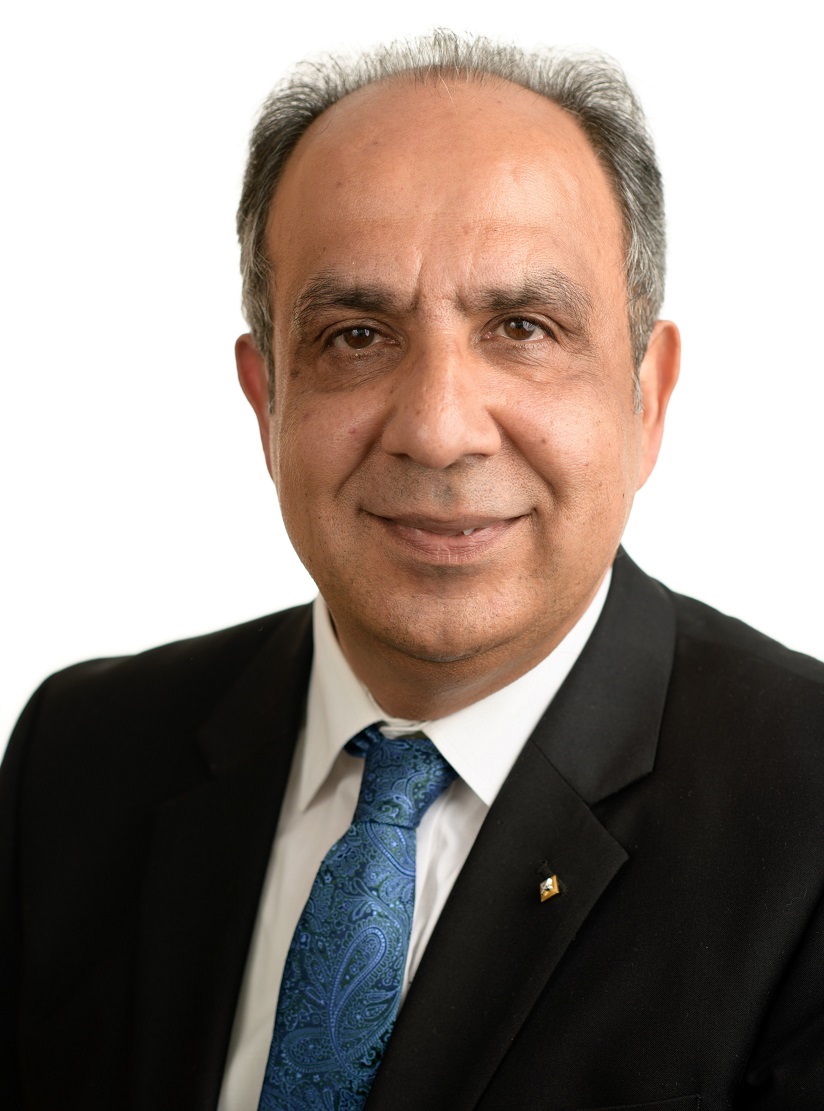}}]{Abbas Jamalipour} (S’86–M’91–SM’00–F’07) holds a PhD in Electrical Engineering from Nagoya University, and he is currently the Professor of Ubiquitous Mobile Networking at the University of Sydney and President of the IEEE Vehicular Technology Society. He is a Fellow of the Institute of Electrical, Information, and Communication Engineers (IEICE) and the Institution of Engineers Australia, an ACM Professional Member, and an IEEE Distinguished Speaker. He has authored nine technical books, eleven book chapters, over 550 technical papers, and five patents, all in the area of wireless communications. Previously, Dr. Jamalipour held the positions of the Executive Vice-President and Editor-in-Chief of VTS Mobile World and has been an elected member of the Board of Governors of the IEEE Vehicular Technology Society since 2014. He was the Editor-in-Chief of IEEE Wireless Communications, Vice President-Conferences, and a member of the Board of Governors of the IEEE Communications Society. He serves as an editor of IEEE Access, IEEE Transactions on Vehicular Technology, and several other journals. He has been a General Chair or Technical Program Chair for a number of conferences, including IEEE ICC, GLOBECOM, WCNC and PIMRC. He is the recipient of a number of prestigious awards, such as the 2019 IEEE ComSoc Distinguished Technical Achievement Award in Green Communications, the 2016 IEEE ComSoc Distinguished Technical Achievement Award in Communications Switching and Routing, the 2010 IEEE ComSoc Harold Sobol Award, the 2006 IEEE ComSoc Best Tutorial Paper Award, as well as 15 Best Paper Awards.
\end{IEEEbiography}

\begin{IEEEbiography}[{\includegraphics[width=1in,height=1.25in,clip,keepaspectratio]{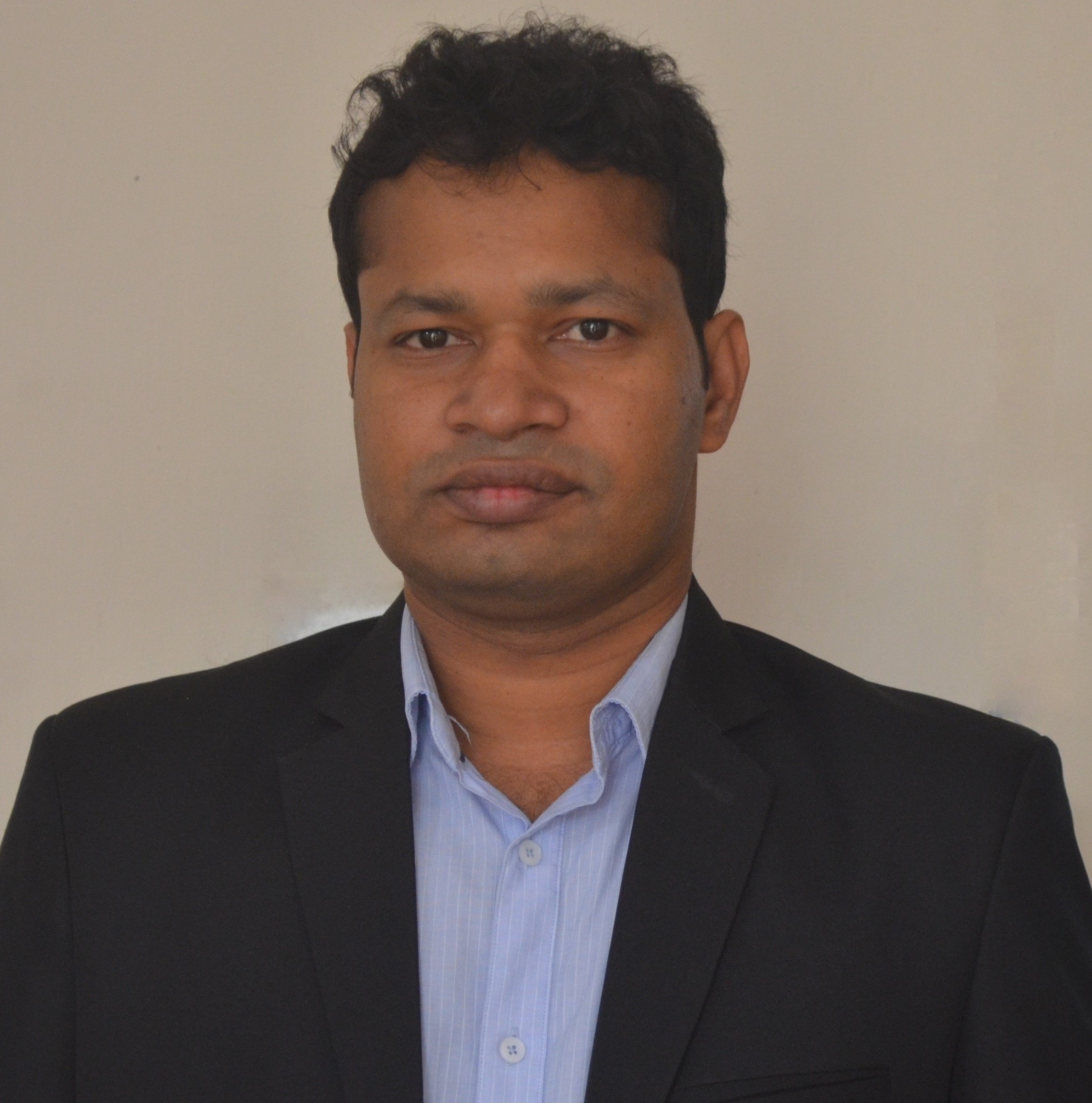}}]{Md. Farhad Hossain } (S’10–M’14) received his Ph.D. from the School of Electrical and Information Engineering of the University of Sydney, Australia in 2014. He completed his B.Sc. and M.Sc. in Electrical and Electronic Engineering (EEE) from Bangladesh University of Engineering and Technology (BUET), Dhaka, Bangladesh in 2003 and 2005, respectively. Currently, he holds a position of Professor in the Department of EEE, BUET. He also works as an electrical and electronic engineering consultant. Dr. Hossain has published over 85 refereed articles in highly prestigious journals and conference proceedings with over 880 citations (h-index: 17). He was the recipient of Best Paper Awards at three international conferences and the Student Travel Grant at the IEEE Global Communications Conference (GLOBECOM), Anaheim, CA, USA, 2012. His research interests include cellular networks, machine learning for wireless networks, smart grid communications, sensor networks, underwater communications, etc. He has been involved in many international journals and conferences in different capacities, including reviewer, TPC member, and organizer.
\end{IEEEbiography}
\EOD

\end{document}